\documentclass[prc,aps,onecolumn,showpacs,preprint]{revtex4}
\usepackage{graphicx}
\usepackage[english,russian]{babel}

\begin{document}

\title{The $K$--matrix approach to the $\Delta -$ resonance
mass--splitting \\
and isospin--violation in low--energy $\pi N$ scattering}

\author{A.~B.~Gridnev$^1$,\ I.~Horn$^2$, \ W.~J.~Briscoe$^3$,\
I.~I.~Strakovsky$^3$}

\affiliation{\rm
$^1$Petersburg Nuclear Physics Institute, Gatchina, Russia.\\
$^2$Helmholtz-Institut f\"ur Strahlen- und Kernphysik,
             Universit\"at  Bonn, Germany\\
$^3$Center for Nuclear Studies, Department of Physics\\
             The George Washington University, U.S.A.}
%\date{\today}

%%%%%%%%%%%%%%%%%%%%%%%%%%%%%%%%%%%%%%%%%%%%%%%%%%%%%
%%%   Abstract
%%%%%%%%%%%%%%%%%%%%%%%%%%%%%%%%%%%%%%%%%%%%%%%%%%%%%
\begin{abstract}

Experimental data on $\pi N$ scattering in the elastic energy
region $T_{\pi} \le$ 250~MeV are analyzed within the multichannel
$K$--matrix approach with effective Lagrangians. Isospin--invariance
is not assumed in this analysis and the physical values for masses
of the involved particles are used.  The corrections due to
$\pi^+-\pi^0$ and $p-n$ mass differences are calculated and found
to be in a reasonable agreement with the NORDITA results.
The results of our analysis describe the experimental observables
very well.  New values for mass and width of the $\Delta^0$ and
$\Delta^{++}$ resonances were obtained from the data.  The
isospin--symmetric version yields phase--shifts values similar
to the new solution, FA02, for the $\pi N$ elastic scattering
amplitude by the GW group based on the latest experimental data.
While our analysis leads to a considerably smaller ($\le $1\%)
isospin--violation in the energy interval T$_{\pi}\sim$30--70~MeV
as compared to 7\% in works by Gibbs \textit{et al}. and Matsinos,
it confirms calculations based on Chiral Perturbation Theory.

\end{abstract}


\maketitle

%%%%%%%%%%%%%%%%%%%%%%%%%%%%%%%%%%%%%%%%%%%%%%%%%%%%%%%%%%%
\section{Introduction}

Low--energy pion--nucleon scattering is one of the fundamental
processes that test the low--energy QCD regime $-$ the pion is a
Goldstone boson in the chiral limit, where the $\pi N$ interaction
goes to zero at zero energy. This behavior is modified by explicit
chiral-symmetry breaking by the small masses of the up and down
quarks, $M_{u}\approx$ 5~MeV and $M_d\approx$ 9~MeV~\cite{leut}.
Since quark masses are not equal, the QCD Lagrangian contains the
isospin--violating term $\propto (M_u-M_d)$.  Calculations using
Chiral Lagrangians~\cite{weinb} and Chiral Perturbation
Theory~\cite{meiss} predict isospin--violation effects for the
low--energy $\pi^{\pm}N$ elastic scattering and charge--exchange
(CEX) reactions $\sim 1$\%. However, for the case of the much
smaller $\pi^0N$ elastic scattering, isospin--breaking is
$\approx$25\%.  Therefore, to observe the isospin--violation
effects, particular experimental conditions are needed, where
these effects are enhanced due to kinematics or other reasons.
One such experiment is found in the $\Delta(1232)$ resonance
mass--splitting measurement. In this case, close to the
$\Delta(1232)$ resonance position, the phase--shifts vary rapidly
with the energy.  Therefore, the small ($\le$1\%) difference
among the masses of the different isospin--states of the
$\Delta(1232)$ resonance as measured in different scattering
channels leads to significant differences for the corresponding
phase--shifts.  The usual procedure for extracting the
$\Delta(1232)$ resonance mass--splitting from the data is in a
phase--shift analysis~\cite{pedroni,koch,abav,bugg,fa02}, where
the $P_{33}$ partial--wave amplitude from $\pi^+p$, $\pi^-p$,
and charge exchange data are considered as independent quantities.
These phase--shifts were then fitted by a Breit--Wigner (BW) formula
to determine the corresponding resonance parameters. The
disadvantage of this procedure is in using isospin--symmetric
quantities in a situation where isospin is not conserved.

Two phenomenological analyses of $\pi N$ scattering data at
low--energies T$_{\pi}\sim$30--70~MeV~\cite{gibbs,mat} reported
about 7\% isospin--violation in the ``triangle relation":

\begin{eqnarray}
f(\pi^-p\to\pi^0n) = \frac{f(\pi^+p) - f(\pi^-p)}{\sqrt{2}}.
\label{1}\end{eqnarray}
This is significantly larger than is predicted in ~\cite{meiss}
and very important for the
determination of the $\Delta(1232)$ resonance mass--splitting,
meaning that an isospin--violation occurs in the background as
well. But this conclusion is based on the rather old and
incomplete experimental data, especially on the charge--exchange
reaction. Note, that the analysis~\cite{gibbs} used preliminary
low--energy $\pi^\pm$ elastic scattering data~\cite{psi}, while in
the final form~\cite{jo}, these data were increased by 10\% in the
absolute values and the pion energies were decreased (shifted) by
1~MeV or more.  In recent years, progress has been made in this
input -- new high--quality experimental data have been published
(see~\cite{isaid} for the up--to--date database). In particular,
detailed experimental data on $\pi^-p\to\pi^0n$ reaction at very
low--energy are reported in~\cite{isenh}. Several years ago, a
$K$--matrix approach with effective Lagrangians was
developed~\cite{goud,gri,feu} and found to be in good agreement
with all $\pi N$ observables in the entire elastic energy region.
In the present paper, we modify this approach to estimate the
isospin--violating effects in the new low--energy, $T_{\pi}\le
250$~MeV, $\pi N$ scattering database.

%%%%%%%%%%%%%%%%%%%%%%%%%%%%%%%%%%%%%%%%%%%%%%
%%%%%%%%%%%%%%%%%%%%%%%%%%%%%%%%%%%%%%%%%%%%%%
\section{Tree--Level Model for the $K$--matrix}

\begin{figure}[ht]
\centering{
\includegraphics[height=0.15\textwidth, angle=0]{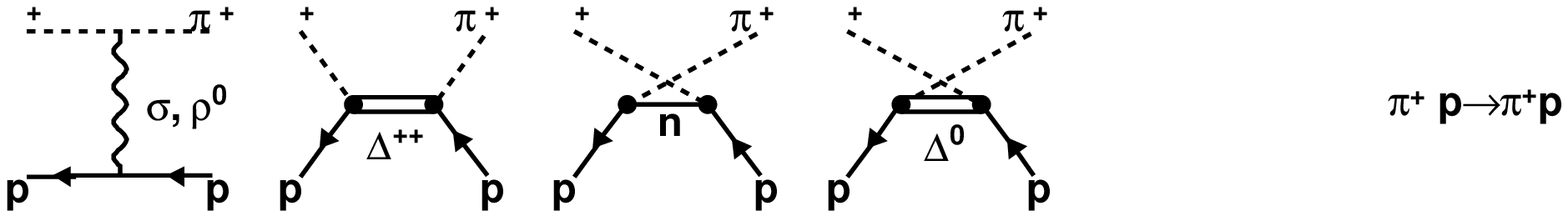}
} \caption{Feynman diagrams for the $\pi^+p$ scattering.
         \label{fig:g1}}
\end{figure}

\begin{figure}[ht]
\centering{
\includegraphics[height=0.4\textwidth, angle=0]{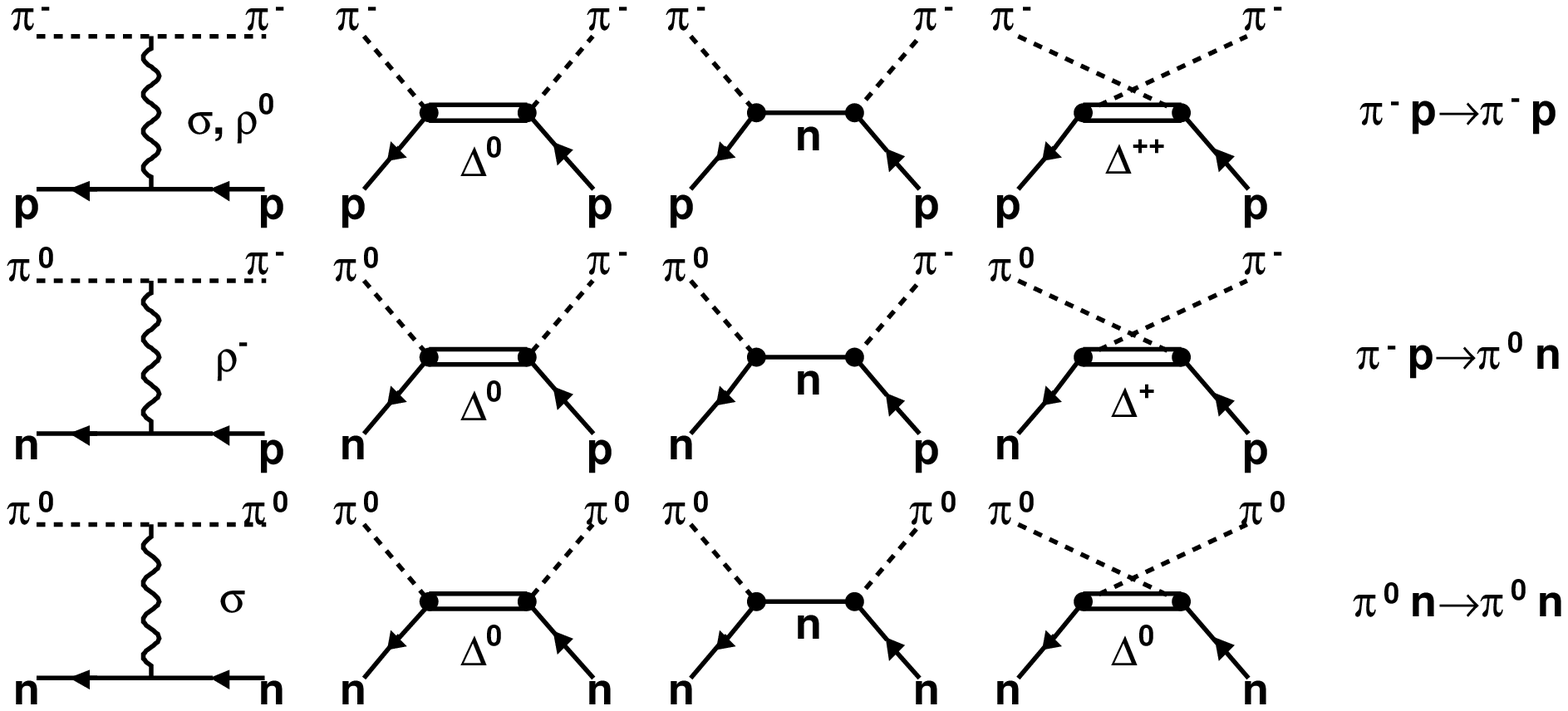}
}
\caption{Feynman diagrams for the $\pi^-p$ scattering.
         \label{fig:g2}}
\end{figure}

The detailed description of the isospin--invariant version of the
multichannel $K$--matrix approach used in this analysis can be found
in~\cite{goud,gri}.  It is assumed that the $K$--matrix, being a
solution of the Bethe--Salpeter equation, can be considered as a
sum of the tree--level Feynman diagrams with the effective
Lagrangians in the vertices. Real part of the loops leads to
renormalization of the mass and couplings. We assume that energy
dependence of the vertex functions in the restricted energy interval
can be accounted for by expansion of these function on power of
invariants.  This leads to Lagrangians, which contain the derivatives
of the fields.  In~\cite{gri}, it was demonstrated that such
approach work very well in isospin--symmetric case up to
$T_{\pi}\sim$~900 MeV.  Here we restrict ourselves the
$\Delta$--resonance energy region $T_{\pi}<$ 250~MeV.  Because we
want to look for possible isospin--violation, we describe $\pi N$
scattering using the same diagrams as in~\cite{goud,gri}, only in the
charged--channels formalism. We confine ourselves to $\pi^{\pm} p$
and $\pi^{0} n$ channels to be able to compare the calculations
with the experimental data. At the hadronic level, the $\pi^{+} p$
scattering is a single--channel problem and the corresponding
Feynman diagrams are shown in Fig.~\ref{fig:g1}. The $\pi^-p$
scattering includes two--channels, one due to the nonzero
$\pi^-p\to\pi^0n$ reaction. Feynman diagrams for this amplitude are
presented in Fig.~\ref{fig:g2}.  The same Lagrangians as in~\cite{goud,gri}
were used in the calculations, but the coupling constants are considered,
in general, to be different for the different channels. The masses of the
incoming and outgoing particles were taken as masses of the physical
particles from PDG~\cite{pdg}. The masses of the intermediate particles
can be different for  different channels as well and for the $\Delta(1232)$
resonance we determine masses of $\Delta^{++}$ and $\Delta^0$ from
the fit of experimental data.  The interaction Lagrangian, corresponding
to the $\pi NN$ vertex, has the structure
\begin{eqnarray}
L_{\pi NN} = -\frac{g_{\pi_{i},N_{k},N_{l}}}{1+x_{i,k,l}}
\overline{\Psi}_{k}
\gamma_5  \left(
i x_{i,k,l} \pi_{i}+\frac{1}{m_{k}+m_{l}}\gamma_\mu \partial^\mu
{ \pi_{i}}
\right) \Psi_{l} + {\rm h.c.},
\label{pinn}\end{eqnarray}
where i,k,l indexes mean pion, left, and right nucleon charged
states. The vertex is assumed to have both pseudoscalor and
pseudovector parts with the mixing parameter $x_{i,k,l}$.

The interaction Lagrangian, corresponding to the $\pi N\Delta$ vertex,
reads as
\begin{equation}
\label{pindelta}
L_{\pi N\Delta} = \frac{g_{\pi_{i},\Delta_{k},N_{l}}}{M_{\Delta_{k}}+m}
\overline{\Psi}_{k}^\mu
\theta_{\mu\nu}T_{kl}
\Psi_{N_{l}} \partial^\nu  \pi_i + {\rm h.c.},
\end{equation}
\begin{equation}
\label{theta_munu}
\theta_{\mu\nu} = g_{\mu\nu} - \left ( Z_{\Delta_{k}} + \frac{1}{2} \right )
\gamma_\mu \gamma_\nu .
\end{equation}
Here, $T_{kl}$ denotes the transition operator between nucleon and
$\Delta$--isobar.We treat the $ \Delta$ -isobar graphs in the most general
manner; thus, constants $Z_{\Delta_{k}}$ and $ g_{\pi_{i},\Delta_{k},N_{l}}$
will be determined
from our fit to the experimental data \\

The interaction Lagrangian for the $\rho\pi\pi$ vertex is put in the form
\begin{eqnarray}
L_{\rho \pi \pi} = -g_{\rho_{i}\pi_{k}\pi_{l}}\rho^{\mu}_{i}
\left (\partial^\mu \pi_{k} \pi_{l} - \partial^\mu \pi_{l} \pi_{k}\right )
\label{rhopipi}\end{eqnarray}
and for the $\rho$NN vertex
\begin{eqnarray}
L_{\rho NN} = -g_{\rho_{i}N_{k}N_{l}} \overline{\Psi}_{k}\frac{1}{2}
\left (\gamma_\mu\rho^{\mu}_{i}+\frac{2\kappa}{m_{k}+m_{l}}\sigma_{\mu
\nu}\partial^\mu \rho^{\nu}_{i}\right )\Psi_{l} + {\rm h.c.}
\label{rhonn}\end{eqnarray}
where $\kappa$ is the tensor and vector coupling constant ratio.
For $\sigma\pi\pi$ and $\sigma NN$ interaction, the following form of
the Lagrangians are used:
\begin{eqnarray}
L_{\sigma \pi \pi} = -g_{\sigma\pi_{i}\pi_{i}}\pi_{i}\pi_{i}\sigma,
\label{sigmapipi}\end{eqnarray}
\begin{eqnarray}
L_{\sigma N N} = -g_{\sigma N_{i} N_{i}}\overline{\Psi}_{i} \Psi_{i}\sigma.
\label{sigmann}\end{eqnarray}

Thus, for isospin--symmetric case, we have seven free parameters:
$G_{\rho}^{V} = \frac{g_{\rho NN}g_{\rho \pi
\pi}}{m_{\rho}^2}$, $G_{\sigma\pi} = \frac{g_{\sigma NN}g_{\sigma \pi
\pi}}{m_{\sigma}^2}$, $\kappa$, $g^{2}_{\pi NN}/4\pi$, $x_{\pi N}$,
$g_{\pi N \Delta}$, and $Z_{\Delta}$.  But futher we assume
different coupling constants for different charged channels, therefore,
the number of free parameters will increase.

The scattering amplitudes can be calculated as:
\begin{eqnarray}
f(\pi^+p\to\pi^+p) = \frac{\tilde{K}_{\pi^+p}}{1 - i\tilde{K}_{\pi^+p}},
\label{2}\end{eqnarray}

\begin{eqnarray}
f(\pi^-p\to\pi^-p) = \frac{\tilde{K}_{\pi^-p} -
i(\tilde{K}_{\pi^-p}\tilde{K}_{\pi^0n} -
\tilde{K}_{\pi^-p\to\pi^0n}^{2})}{\tilde{K}_{\pi^-p}\tilde{K}_{\pi^0n} -
\tilde{K}_{\pi^-p\to\pi^0n}^{2} - i(\tilde{K}_{\pi^-p}+\tilde{K}_{\pi^0n})},
\label{3}\end{eqnarray}

\begin{eqnarray}
f(\pi^-p\to\pi^0n) =
\frac{\tilde{K}_{\pi^-p\to\pi^0n}}{\tilde{K}_{\pi^-p}\tilde{K}_{\pi^0n} -
\tilde{K}_{\pi^-p\to\pi^{0} n}^{2} -
i(\tilde{K}_{\pi^-p}+\tilde{K}_{\pi^0n})},
\label{4}\end{eqnarray}

\begin{eqnarray}
f(\pi^0n\to\pi^0n) = \frac{\tilde{K}_{\pi^0n} -
i(\tilde{K}_{\pi^-p}\tilde{K}_{\pi^0n} -
\tilde{K}_{\pi^-p\to\pi^0n}^{2})}{\tilde{K}_{\pi^-p}\tilde{K}_{\pi^0n} -
\tilde{K}_{\pi^-p\to\pi^0n}^{2} -
i(\tilde{K}_{\pi^-p}+\tilde{K}_{\pi^0n})},
\label{5}\end{eqnarray}
where, for example, $\tilde{K}_{\pi^0n}$ is the $K$--matrix element for
$\pi^0n\to\pi^0n$ channel multiplied by c.m. momentum of the
$\pi^0n$ system to obtain the dimensionless scattering
amplitudes in Eqs.~(\ref{2}--\ref{5}).
%%%%%%%%%%%%%%%%%%%%%%%%%%%%%%%%%%%%%%%%%%%%%
\section{Electromagnetic corrections}

In the analysis of the pion--nucleon experimental data, the ~$SU(2)$
isospin--symmetry is usually assumed. This implies that the masses
of the isospin--multiplet must be equal.  However, the physical
masses of the particles are different. There are two sources for
this mass--splitting: the electromagnetic self--energy and the QCD
quark mass difference. The latter leads to isospin--violating
effects in the strong interaction. Usually the influence of the
mass difference on the pion--nucleon scattering amplitude is
calculated together with the true electromagnetic corrections. The
most popular way to do this is the method of the effective
potential~\cite{gashi} and dispersion relations~\cite{trom}.

There are several reasons for using $K$--matrix approach to
calculate the mass difference corrections (MDC) to the
pion--nucleon amplitude:

\begin{itemize}
\item Only the mass--splitting of the external particles (pions
      and nucleons) have been taken into account up to now.
      In the $K$--matrix approach, corrections due to mass
      difference of all particles in the intermediate
      state can be calculated explicitly.
\item MDC contributions dominate the structure of the total
      $P$--waves corrections~\cite{trom}. Therefore, it is
      important to estimate them by different methods to
      obtain reliable results.
\end{itemize}

Let us start with  $\pi^-p$ elastic scattering.  In general, the
$S$--matrix for the charged channels has the form:

\begin{eqnarray}
    S_c = \left(
\begin{array}{cccc}
S_{\pi^-p}& ; &S_{\pi^-p\to\pi^0n}\\
S_{\pi^0n\to\pi^-p}&; &S_{\pi^0n}
\end{array}
    \right).
\label{6}\end{eqnarray}

Time--reversal invariance is assumed, thus, $S_{\pi^-p\to\pi^0n}$ =
$S_{\pi^0n\to\pi^-p}$. If ~$SU(2)$ isospin--symmetry is valid, then
the unitarity matrix

\begin{eqnarray}
U_t = \frac{1}{\sqrt{3}}\left(
\begin{array}{cccc}
 -\sqrt{2}& ; &1\\
    1&; &\sqrt{2}
\end{array}
    \right)
\label{7}\end{eqnarray}
can be used for the transformation of the $S_c$ to the isospin--basis

\begin{eqnarray}
S_I = U_t^+S_cU_t.
\label{8}\end{eqnarray}

In this case, the isospin--channels are eigenchannels, therefore,
$S_I$ must be diagonal:

\begin{eqnarray}
S_I^H = \left(
\begin{array}{cccc}
 \eta_{H1}e^{2i\delta_{H1}}& ; &0\\
    0&; &\eta_{H3}e^{2i\delta_{H3}}
\end{array}
    \right).
\label{9}\end{eqnarray}
Here the index H means that quantities are calculated for eigenchannels.
If isospin--symmetry is slightly violated, then the transformation (\ref{8})
leads to nondiagonal matrix elements.  To be able to compare the results
with~\cite{trom}, we write the $S$--matrix for this case in the following
form:

\begin{eqnarray}
    S_{I} = \left(
\begin{array}{cccc}
\eta_1e^{2i\delta_1}& ; &\frac{2}{3}\sqrt{2}(\eta_{13}+i\Delta_{13})
e^{i(\delta_1+\delta_3)}\\
\frac{2}{3}\sqrt{2}(\eta_{13}+i\Delta_{13})e^{i(\delta_1+\delta_3)}&;
&\eta_3e^{2i\delta_3}
\end{array}
    \right).
\label{10}\end{eqnarray}
Then, the corrections to the isospin--symmetric matrix $S_I^H$, apart
from $\Delta_{13}$ and $\eta_{13}$ are defined as:

\begin{eqnarray}
   \delta_1 =
   \delta_{H1} - \frac{2}{3}\Delta_1 ~ ;  ~
   \bar\eta_1 = \eta_{H1} - \eta_1; \nonumber \\
   \delta_3 = \delta_{H3} - \frac{1}{3}\Delta_3 ~; ~
   \bar\eta_3 = \eta_{H3} - \eta_3.
\label{11}\end{eqnarray}

There are different contributions to these corrections: due to
electromagnetic interactions, mass--splitting, $n\gamma$ channel,
\textit{etc}. Here, we are only interested in the mass--splitting
contribution. In order to calculate it, the isospin--symmetric
reference masses have to be fixed. The conventional choices are $m
= M_p$ for the nucleon, $M_{\pi} = M_{\pi^{\pm}}$ for the charged
pions, and $M_{\Delta}$ = 1232~MeV is added for the $\Delta(1232)$
resonance. The procedure of the calculations is as follows. First
we use the $K$--matrix  with the physical masses to calculate the
$S$--matrix for the charged channels $S_c$.  Then, we transform $S_c$
to isospin--basis by matrix (\ref{7}) and obtain the corrections
$\Delta_{13},\eta_{13}$ and the quantities $\eta_1,\delta_1,\eta_3$,
and $\delta_3$. After that, we repeat the calculations with the
isospin--symmetric reference masses and obtain the values $\eta_{H1},
\delta_{H1}, \eta_{H3}$, and $\delta_{H3}$. Finally, the corrections
$\bar\eta_1,\Delta_1,\bar\eta_3$, and $\Delta_3$ are found.

For the $\pi^{+} p$ scattering, we use the one--channel form for
the nuclear $S_{\pi^+p}$ and the hadronic $S_H$ matrices (see~\cite{trom}
for the definition of the terms):

\begin{eqnarray}
S_{\pi^+p} = \eta_3^+e^{2i\delta_3^+} ~;~
S_H = \eta_{H3}^+e^{2i\delta_{H3}^+},
\label{12}\end{eqnarray}
and the similar formulae for the corrections:

\begin{eqnarray}
\delta_3^+ = \delta_{H3}^+ - \frac{1}{3}\Delta_3^+ ~ ;  ~
\bar\eta_3^+ = \eta_{H3}^+ - \eta_{3}^+.
\label{13}\end{eqnarray}
We find that contribution of the mass--splitting effect on the
inelasticity corrections $\bar\eta_1,\bar\eta_3$, and
$\bar\eta_3^+$ is very small (less than $10^{-4}$) and can be
neglected.

\begin{figure}[ht]
\centering{
\includegraphics[height=0.45\textwidth, angle=90]{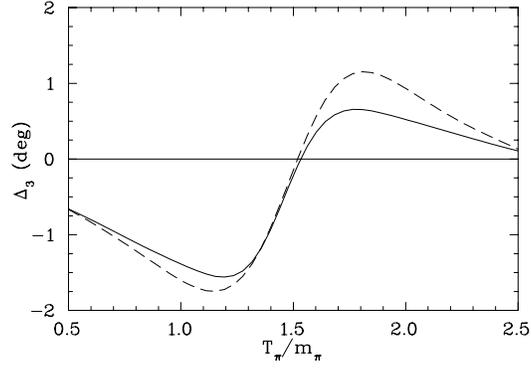}
}
\caption{The $\Delta_{3}$ mass correction.  Solid and dashed
         lines represent results of the present
         (NORDITA~\protect\cite{trom}) work.
         \label{fig:g3}}
\end{figure}

In Fig.~\ref{fig:g3}, the most important $\Delta_{3}$
mass difference correction for total angular momentum
J = 3/2 is presented (solid line) together with that from
~\cite{trom} (dashed line).  The $\Delta(1232)$
resonance mass--splitting is not taken into account in
this figure, as it was in~\cite{trom}.  But it was found
that $\Delta(1232)$ resonance mass--splitting leads to
very large effect for the $\Delta_{3}$ correction (this
is the correction to $P_{33}$ phase--shifts).  The origin
of that comes from the rapid variation of phase shift
near the resonance position.  Therefore, the small change
in the $\Delta(1232)$ mass corresponds to a large phase
shift difference. This $\Delta(1232)$ resonance
mass--splitting effect is not contained in the NORDITA
~\cite{trom} corrections.  Therefore, if one wants to
extract hadronic isospin--symmetric amplitudes, based on
the NORDITA procedure, one has to include the $\Delta(1232)$
mass--splitting effect. If $\Delta(1232)$ mass--splitting
effect is not included, then $P_{33}$ phase--shifts
determined from $\pi^+p$ and $\pi^-p$ will be
different~\cite{trom,abav}.  These results for other
partial--waves are small and of the same order as true
electromagnetic ones.

%%%%%%%%%%%%%%%%%%%%%%%%%%%%%%%%%%%%%%%%%%%%%%%
\section{Database and fitting procedure}

For a definite set of the coupling constants and particle masses,
the hadronic part of the amplitude was calculated according to
graphs in Figs.~\ref{fig:g1} and \ref{fig:g2}. The electromagnetic
interaction was added in order to compare with experimental data.
We use the observed masses of the particles, therefore, the
electromagnetic parts of NORDITA corrections were included only
using isospin--invariance relations. It was found, however, that the
latter do not affect the values of the extracted parameters within
the uncertainties and can be neglected.  In order to determine the
parameters of the model, the standard MINUIT CERN library
program~\cite{minuit} was used.  The experimental data used here
are those $\pi N$ data which can be found in the SAID
database~\cite{isaid}. In the present work, we confine ourselves
to partial--waves with spin 1/2 and 3/2. Only for these can the
Lagrangians be written in ``conventional" way (see discussion
on Lagrangian for spin 3/2 particles in~\cite{nieu}). The
inelastic channels are not included in the present version of the
model; therefore, only the data below T$_{\pi}$ = 250~MeV are used
in the fit. In this energy region, there are no open inelastic
channels and the $S$ and $P$ partial--waves give the dominant
contribution to the observables. The small contributions of the
higher partial--waves were taken from the partial--wave analyses.
The different partial--wave analyses (KH80~\cite{koch},
KA84~\cite{ka84}, KA85~\cite{ka85}, SM95~\cite{sm95}, and
FA02~\cite{fa02}) lead to similar results, and only the latest FA02
solution  was used in all further calculations.  As a rule, the
parameter values obtained by fit have very small uncertainties.
Therefore, the main source of these uncertainties comes from the
database.  To estimate it, we perform the fit in two steps.
First, we take all data in the first fit.  Then remove the data
points which give more than 4 in $\chi^2$ units (mainly
from~\cite{jo,br})and perform a second fit.  The number of such
points is about 2\%.  We take the difference in the parameter
values in these two fits as the uncertainties of the parameter. It
was found that rejection of more data leads to parameter values
within the uncertainties determined above.  Typically, a
$\chi^2\sim 1.5$ was obtained. The largest contribution to
$\chi^2$ comes from $\pi^-p$ elastic scattering data.\\

%%%%%%%%%%%%%%%%%%%%%%%%%%%%%%%%%%%%%%%%%%%%%%%%%%%%%%%
\section{Results and discussion}

As a first step, and in order to reduce the number of the free
parameters, we assume the coupling constant entering the
interaction Lagrangians to be isospin--invariant. Thus, we have
nine free parameters -- seven for coupling constants~\cite{goud}
and two for
$M_{\Delta^{++}}$ and $M_{\Delta^0}$. The results for the coupling
constants were found to be similar to those from~\cite{goud}
and are presented in the Table~\ref{tbl1}.  The value of $g_{\pi
NN}^{2}/4\pi$ = 13.80 agrees very well with the recent
result $g^{2}_{\pi NN}/4\pi$ = 13.75 $\pm$ 0.10 from FA02
solution~\cite{fa02}.

In order to determine the resonance parameters, we should
assume some procedure to separate the resonance and
background contributions to the amplitude.  In general,
such a procedure is somewhat arbitrary~\cite{woolc}. The
most popular way is to write the BW formula for the
scattering amplitude near the resonance position. For the
one--channel case ($\pi^+p$ scattering), this corresponds to
defining the mass of the resonance as the pole position
of the corresponding $K$--matrix. Indeed, close to the pole
the $K$--matrix has a form:

\begin{eqnarray}
K(w) = \frac{\alpha^2}{w-M} + \beta(w) ,
\label{14}\end{eqnarray}
where $\alpha$ and $\beta(w)$ are a smooth functions of
the energy.  Then in this region the scattering amplitude
obtains the BW form:
\begin{eqnarray}
F(w) = \frac{K}{1-iqK} = \frac{\alpha^2 +
\beta(w)(w-M)}{w-M+iq[\alpha^2 +
\beta(w)(w-M)]}\approx\frac{\alpha^2}{w-M+iq\alpha^2} .
\label{15}\end{eqnarray}
Therefore, the width of the resonance is read as
\begin{eqnarray}
\Gamma_{\Delta^{++}} = 2\lim_{w \to M_{\Delta^{++}}}(w -
M_{\Delta^{++}})~q_{\pi^+}K_{\pi^+p} .
\label{16}\end{eqnarray}

For the two--channel case ($\pi^-$p scattering), the form of
amplitudes (\ref{3}--\ref{5}) is more complicated. But the
situation can be improved using the eigenchannel representation.
This means that we define the new channel basis to transform the
$K$--matrix into the diagonal form K$^{ech}$ = $U^+ K U$, where $U$
is the unitary transformation matrix.  At the same time, the
matrix of amplitudes also becomes diagonal.  Only one channel
contains the resonance in this representation (see Appendix in
~\cite{gri} for details) with the same pole position. For this
channel, the amplitude has a BW form as in the one--channel case.
In order to calculate the width of the resonance, trace
conservation under unitarity transformations $tr(K^{ech}) = tr(K)$
is used.  Therefore:
\begin{eqnarray}
   \Gamma_{\Delta^0}& = &2\lim_{w\to M_{\Delta^0}}(w -
M_{\Delta^0})tr(K^{ech}) = 2\lim_{w\to M_{\Delta^0}}(w -
M_{\Delta^0})tr(K) \nonumber \\
   & = &2\lim_{w\to M_{\Delta^0}}(w -
   M_{\Delta^0})\left(q_{\pi^-}K_{\pi^-p} + q_{\pi^0}K_{\pi^0n}\right) .
\label{17}\end{eqnarray}

In order to clarify the procedure, let us consider as an
example the $\pi^-p$ scattering, when the isospin is
conserved.  The charged channels are $\pi^-p\to\pi^-p$
and $\pi^-p\to\pi^0n$.  So, the scattering amplitude is a
$2\times2$ matrix.  After transforming this matrix from
the charged--channel basis to isotopic one, we get a
$2\times2$ diagonal matrix.  Now only one channel with
isospin 3/2 contains the resonance ($\Delta(1232)$).

Our fitting procedure leads to reasonable values for
masses and widths of the $\Delta(1232)$ isobar; these are
presented in the Table~\ref{tbl2} together with PDG data.
It should be noted that all values from PDG were not
obtained directly from the data as in the present work,
but by using the results of the phase shift analyses for
individual charged channels. Then, the resulting
amplitudes (or remaining part of data as in ~\cite{fa02})
are fitted by some simple BW formula. The graphs in
Figs.~\ref{fig:g1} and \ref{fig:g2} with
$\Delta(1232)$ resonance in the intermediate state give
contributions not to $P_{33}$ partial--wave only, but to
all other waves too.  As a result, in the resonance
region, we found the 1\% isospin--violation in $S_{31}$
and somewhat smaller in $P_{31}$ partial--waves due to
difference in the $\Delta(1232)$ masses. This was not
accounted for in above procedure, but it is important in
determining the $\Delta(1232)$ width because of the
rather wide energy interval used in the fit. This can
account for the large spread of
$\Gamma_{\Delta^0}-\Gamma_{\Delta^{++}}$ values in the
Table~\ref{tbl2}. In our approach, the fit to different
data combinations ($(\pi^+,\pi^-),~(\pi^+,CEX)$, and
$(\pi^-,CEX)$) gives the same values for all parameters
with slightly larger uncertainties.

\begin{figure}[ht]
\centering{
\includegraphics[height=0.45\textwidth, angle=90]{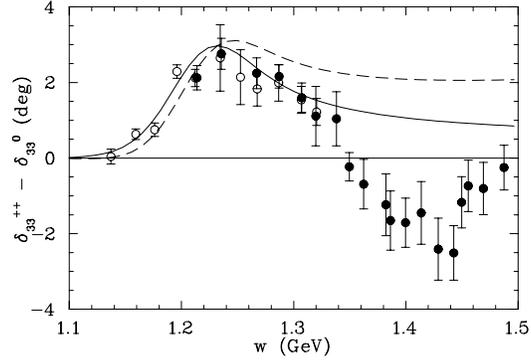}
}
\caption{Energy--dependence of the phase--shifts difference
          $\delta^{++}_{33}(w)-\delta^{0}_{33}(w)$.
          Solid line shows the result of the present
          work. The dashed line corresponds to the
          case, when $g_{\pi^+p\Delta}$ is increased by 1\%
          in comparison with the corresponding couplings
          for the other charged channels.  Filled
          ~\protect\cite{abav} and open
          ~\protect\cite{bugg} circles represent
          results of previous partial--wave analyses.
          \label{fig:g4}}
\end{figure}

\begin{figure}[ht]
\centering{
\includegraphics[height=0.45\textwidth, angle=90]{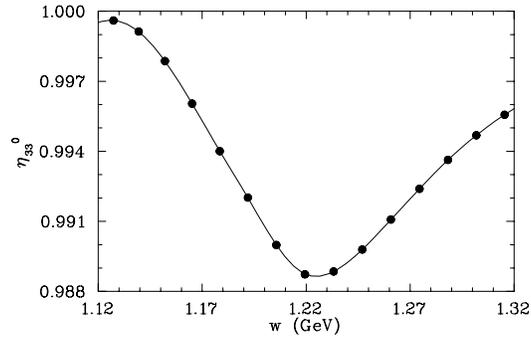}
}
\caption{Energy dependence of the inelasticity parameter
         $\eta^0_{33}(w)$.  Solid line shows result of
         the present work.  Filled circles represent
         results (the uncertainties are within the symbols)
         from~\protect\cite{bugg}.
         \label{fig:g5}}
\end{figure}

\begin{figure}[ht]
\centering{
\includegraphics[height=0.4\textwidth, angle=90]{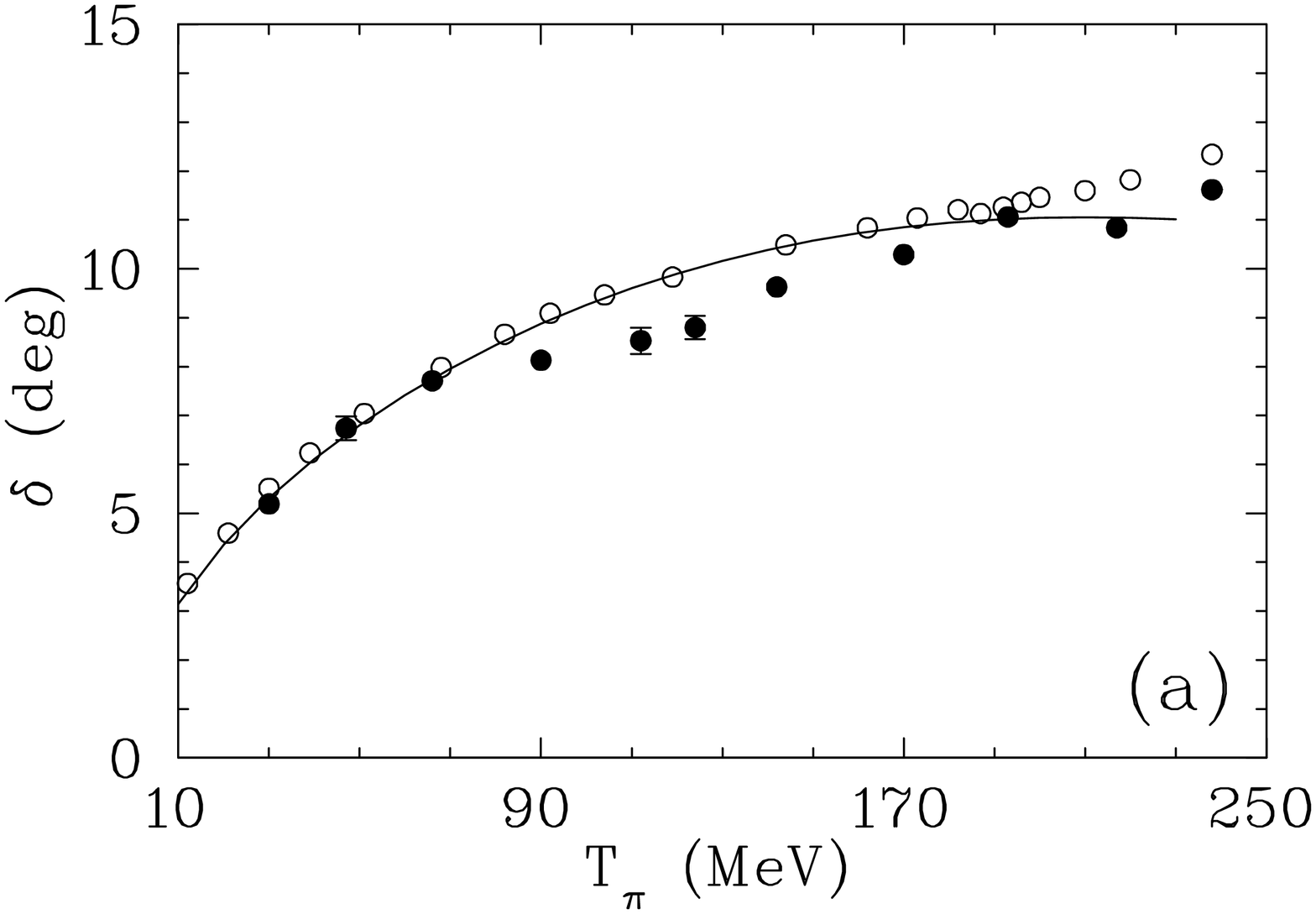}\hfill
\includegraphics[height=0.4\textwidth, angle=90]{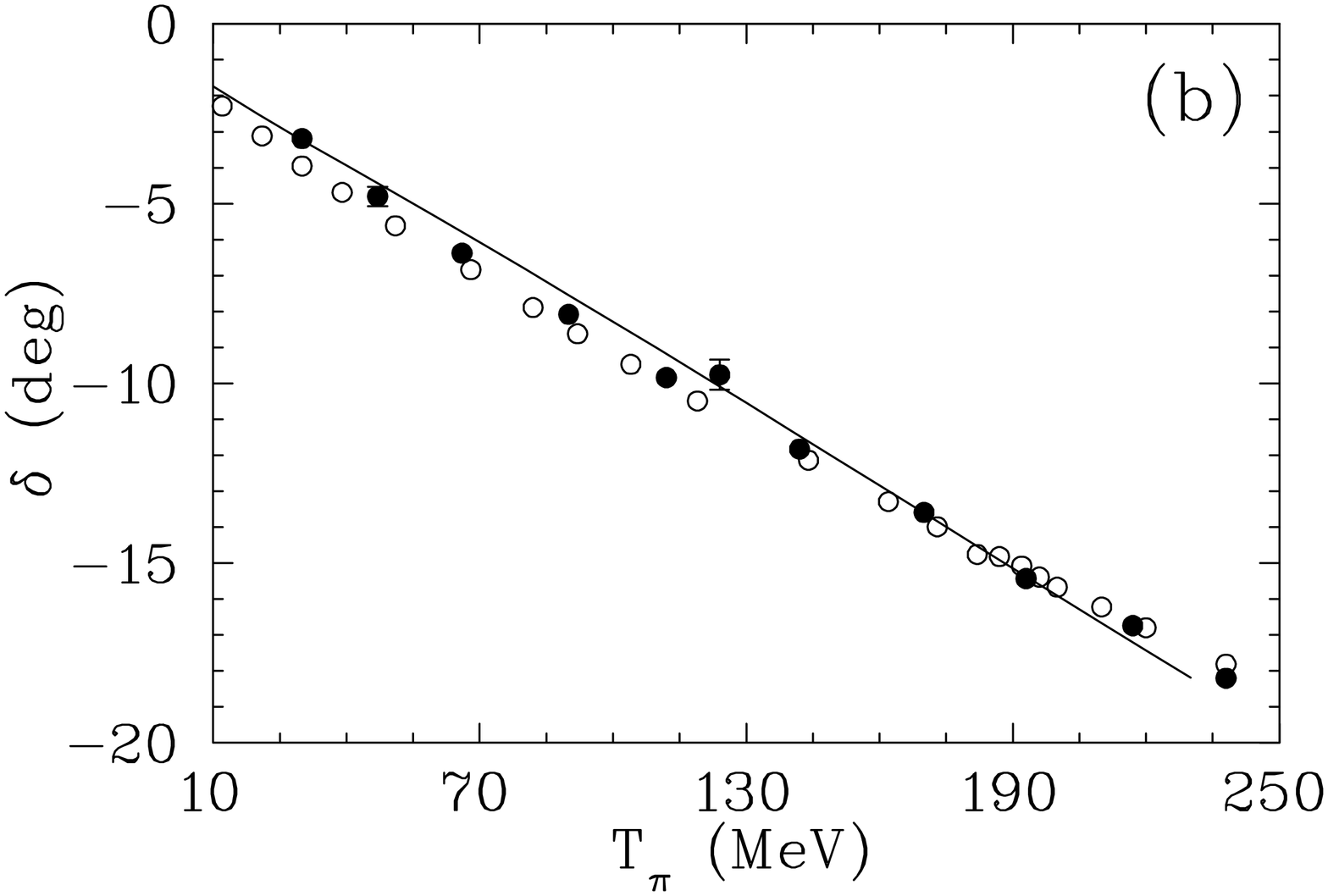}
\includegraphics[height=0.4\textwidth, angle=90]{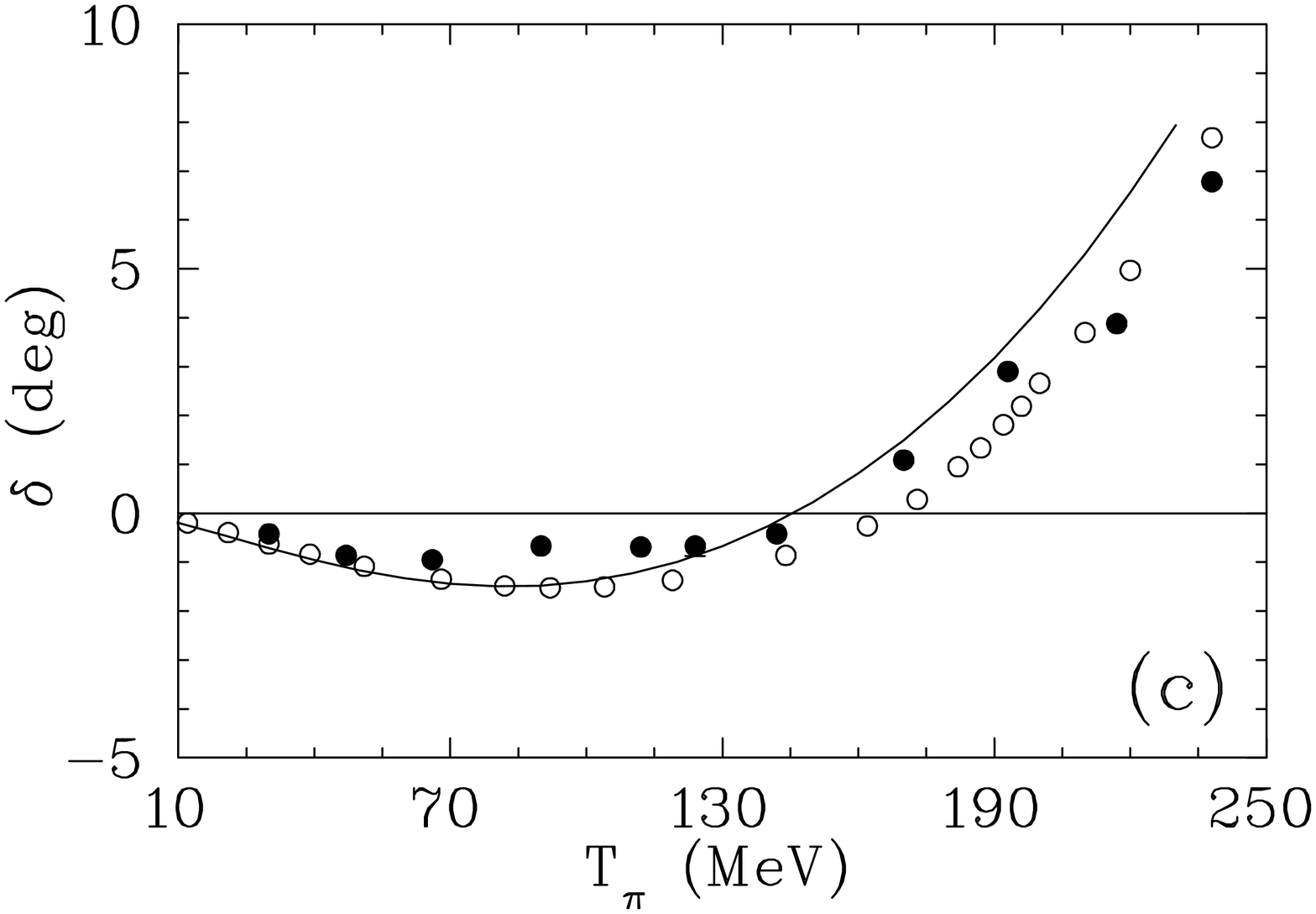}\hfill
\includegraphics[height=0.4\textwidth, angle=90]{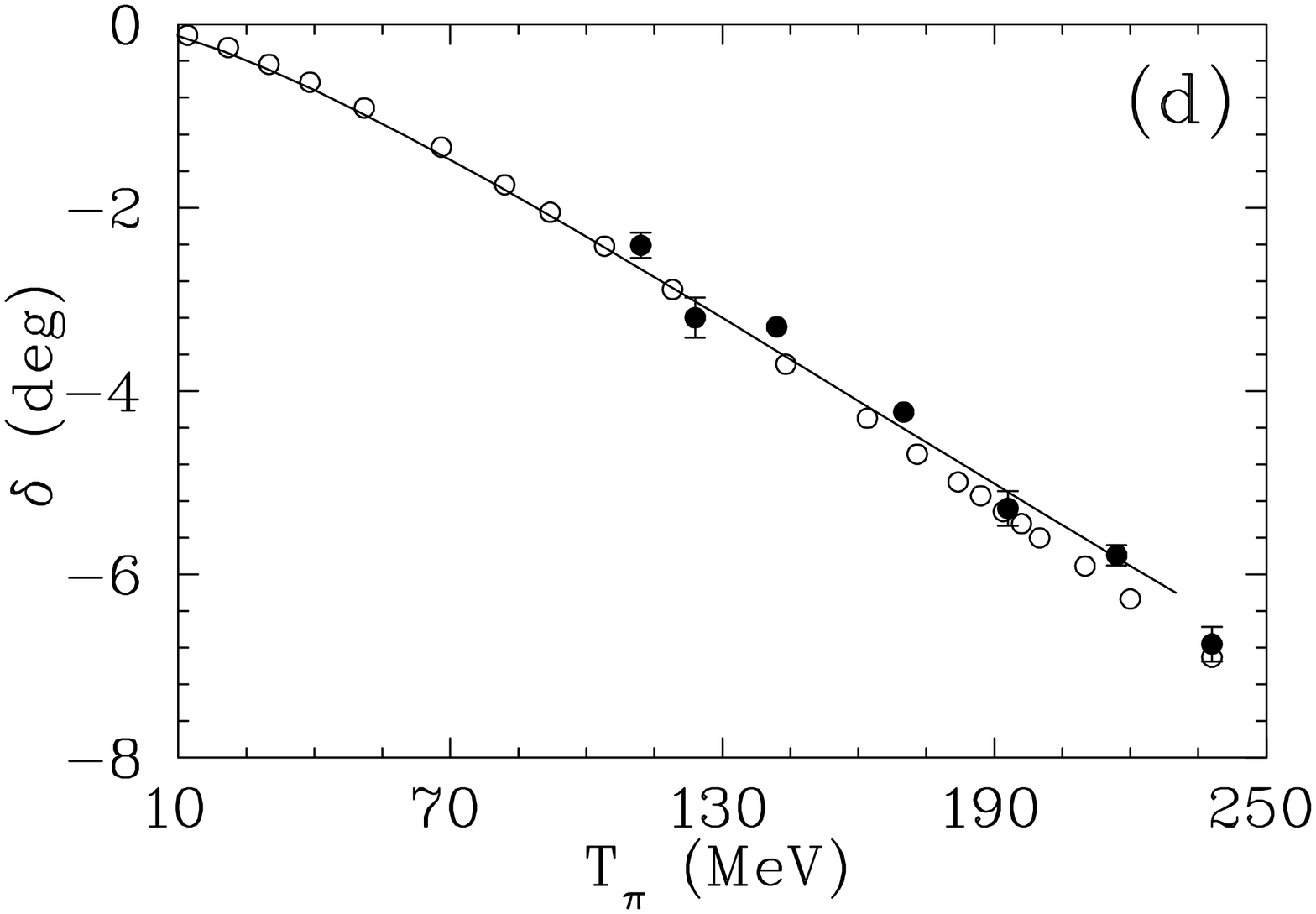}
\includegraphics[height=0.4\textwidth, angle=90]{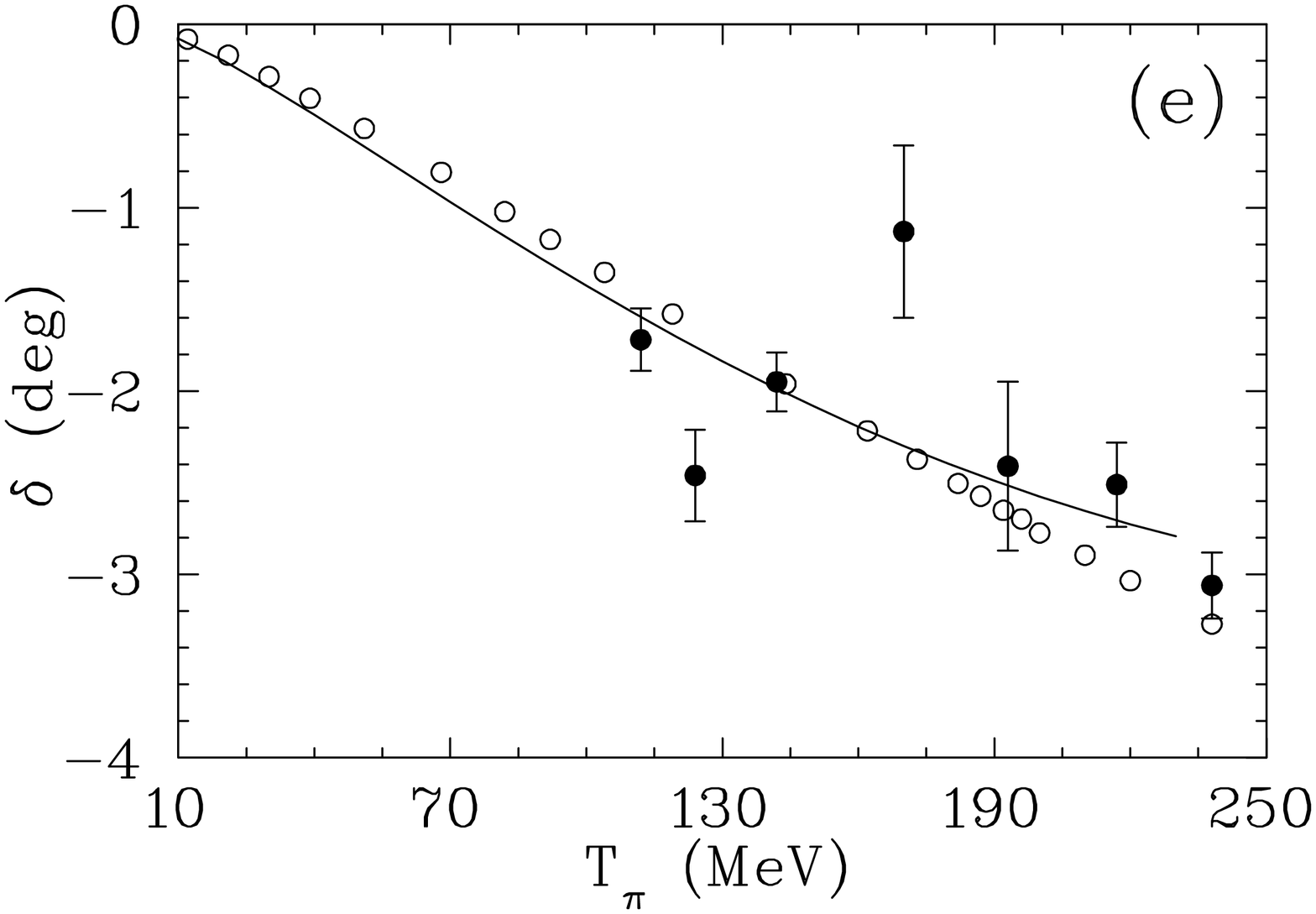}\hfill
\includegraphics[height=0.4\textwidth, angle=90]{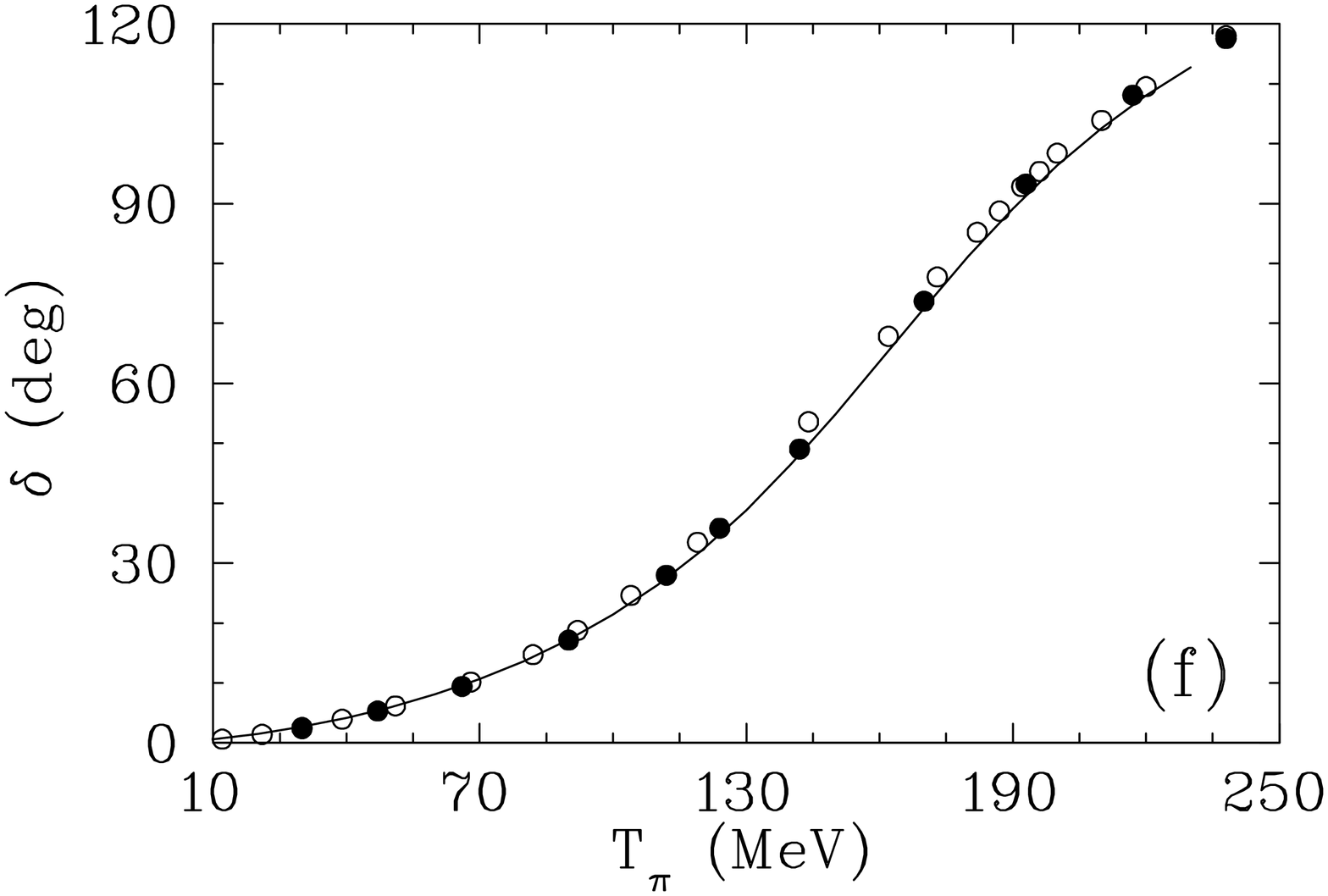}
}
\caption{The energy dependence of the $S$-- and $P$--
         phase--shifts.  Solid lines show results of the
         present work.  (a) $S_{11}$, (b) $S_{31}$,
         (c) $P_{11}$, (d) $P_{31}$, (e) $P_{13}$,
         and (f) $P_{33}$.  The solid (open) circles
         are the GW SAID single--energy solutions
         associated with FA02~\protect\cite{fa02}
         (Karlsruhe KH80~\protect\cite{koch}, the
         uncertainties are within the symbols).
         \label{fig:g6}}
\end{figure}

We obtain equal coupling constants in all charged channels
(see below), therefore, the difference in the $\Delta(1232)$
widths has two sources: the difference in phase space due to
different $\Delta(1232)$ masses (this gives 3.7~MeV) and
different masses of final particles in the $\Delta^0\to\pi^0n$
and $\Delta^{++}\to\pi^+p$ decays (this gives 0.9~MeV).  We
obtain $\Gamma_{\Delta^0\to\pi^0n}/\Gamma_{\Delta^0\to\pi^-p}$
= 2.024 instead of 2.0 for the isospin--invariant case. There
is also an additional $\sim$ 1~MeV contribution to the $\Delta^0$
width from the $\Delta^0\to\gamma n$ decay, which is not included
in present version of the model. In Refs.~\cite{abav,bugg}, the
quantities $\delta^{++}_{33}(w), ~\delta^{0}_{33}(w)$, and
$\eta^{0}_{33}(w)$ via

\begin{eqnarray}
f^{3/2}_{\pi^+p}(w) = \frac{e^{2i\delta^{++}_{33}(w)}-1}{2iq_{\pi^+}} ,
\label{18}\end{eqnarray}

\begin{eqnarray}
f^{3/2}_{\pi^-p}(w)+\sqrt{2}f^{3/2}_{\pi^-p\to\pi^0n}(w) =
\frac{\eta^0_{33}(w)e^{2i\delta^0_{33}(w)}-1}{2iq_{\pi^-}}
\label{19}\end{eqnarray}
\\
were determined from the experimental data. In Figs.~\ref{fig:g4}
and \ref{fig:g5}, we compare calculated values for
$\delta^{++}_{33}(w)-\delta^{0}_{33}(w)$ and ~$\eta^{0}_{33}(w)$
with the results of Refs.~\cite{abav,bugg}.  As it is seen from
the figures, the agreement is very good up to $W\approx$1.3~GeV.
As was found in~\cite{abav} at higher energies, the
difference $\delta^{++}_{33}(w)-\delta^0_{33}(w)$ changes sign.
Such behavior cannot be explained within our approach.  However,
just at these energies an inelastic two--pion production process
opens; the difference in the pion masses is a probable source of
this phenomena. In Fig.~\ref{fig:g6}, the phase--shifts for the
isospin--symmetric case (masses of all particles were set to
conventional values) are shown together with the results of
phase--shift analyses KH80 and FA02 -- our results are not in
conflict with the known phase shift analyses.
In~\cite{fett} a large discrepancy between calculations based on
Chiral Perturbation Theory and results of phase shift analysis in
$S$--waves was found for $P_{\pi} <$ 100~MeV/c. From Fig.~\ref{fig:g6}
we see that there is no such a discrepancy in our approach.

As a next step, we tried to allow some coupling constants to be
different for different charged channel.  No statistically proved
differences in the $g_{\pi^+NN}, g_{\pi^-NN}$, and $g_{\pi^0NN}$
or other coupling constants were found. It is interesting to
note that all fits give nearly equal values of $g_{\pi N \Delta}$
for all charged channels with very small uncertaintie, less then 0.2\%. In
Fig.~\ref{fig:g4}, the dashed line shows the results for
$\delta^{++}_{33}(w)-\delta^{0}_{33}(w)$, when we increase
$g_{\pi^+p\Delta}$ by 1\% in comparison with the corresponding
couplings for the other charged channels.

To look for another source for isospin--breaking, we add the
$\rho\omega$ mixing to the $K$--matrix as in~\cite{bg} but
allow the mixing parameter $H_{\rho\omega} $ to be
free. The data show no evidence for such mechanism and the fit
gives nearly a zero value for $H_{\rho\omega} $.

Thus, we did not find any isospin--breaking effects except that due
to the $\Delta$ mass difference. However, in Refs.~\cite{gibbs,mat}
a 7\% violation of ``triangle relation" was found in the analysis
of the same data within the T$_{\pi}\sim$30--70~MeV energy region.
Therefore, following these works, we performed a fit to CEX data
alone and then compared it with the results of the combined fit of
the $\pi^+p$ and $\pi^-p$ elastic scattering data. We now look at
the results for the $S$--wave part $f^s$ of the scattering amplitude
at $T_{\pi} = 30$~MeV. From the fit of the CEX data alone, we
obtain $f^s_{CEX}$ = -0.1751~fm, whereas from the combined fit
$f^s_{\pi^+,\pi^-}$ = -0.1624~fm , which implies a 7\% violation of
the ``triangle inequality". This violation cannot be explained by
$\Delta(1232)$ mass difference alone.  The possible reason for such
a discrepancy is the procedure itself. The $\pi^-p $ elastic and
CEX are coupled channels even if isospin is not conserved. Therefore,
some changes in the CEX amplitude should lead to corresponding
changes in the $\pi^-p$ elastic scattering amplitude. This means
that we cannot fit CEX data separate from the $\pi^-p$ elastic data
or inconsistent results could be obtained. To demonstrate this, we
perform the individual fits to $\pi^-p$ and $\pi^+p$ elastic data.
The results are: $f^s_{\pi^+}$ = -0.1397~fm and $f^s_{\pi^-}$ =
0.1020~fm. These values lead to 2.4\% violation of ``triangle
relation" only. This demonstrates that the above procedure is
somewhat indefinite. The only way to check for isospin--violation
is to compare the results of the combined fit of $\pi^-p$ elastic
and CEX data with the corresponding quantity from $\pi^+p$ data.
Doing this, we obtain $f^s_{\pi^+}$ = -0.1376~fm, which is in good
agreement with $f^s_{\pi^+}$ = -0.1397~fm from $\pi^+p $ data
alone, taking into account the $\approx $ 1.0\% uncertaintie in the
amplitude.

\begin{figure}[ht]
\centering{
\includegraphics[height=0.4\textwidth, angle=90]{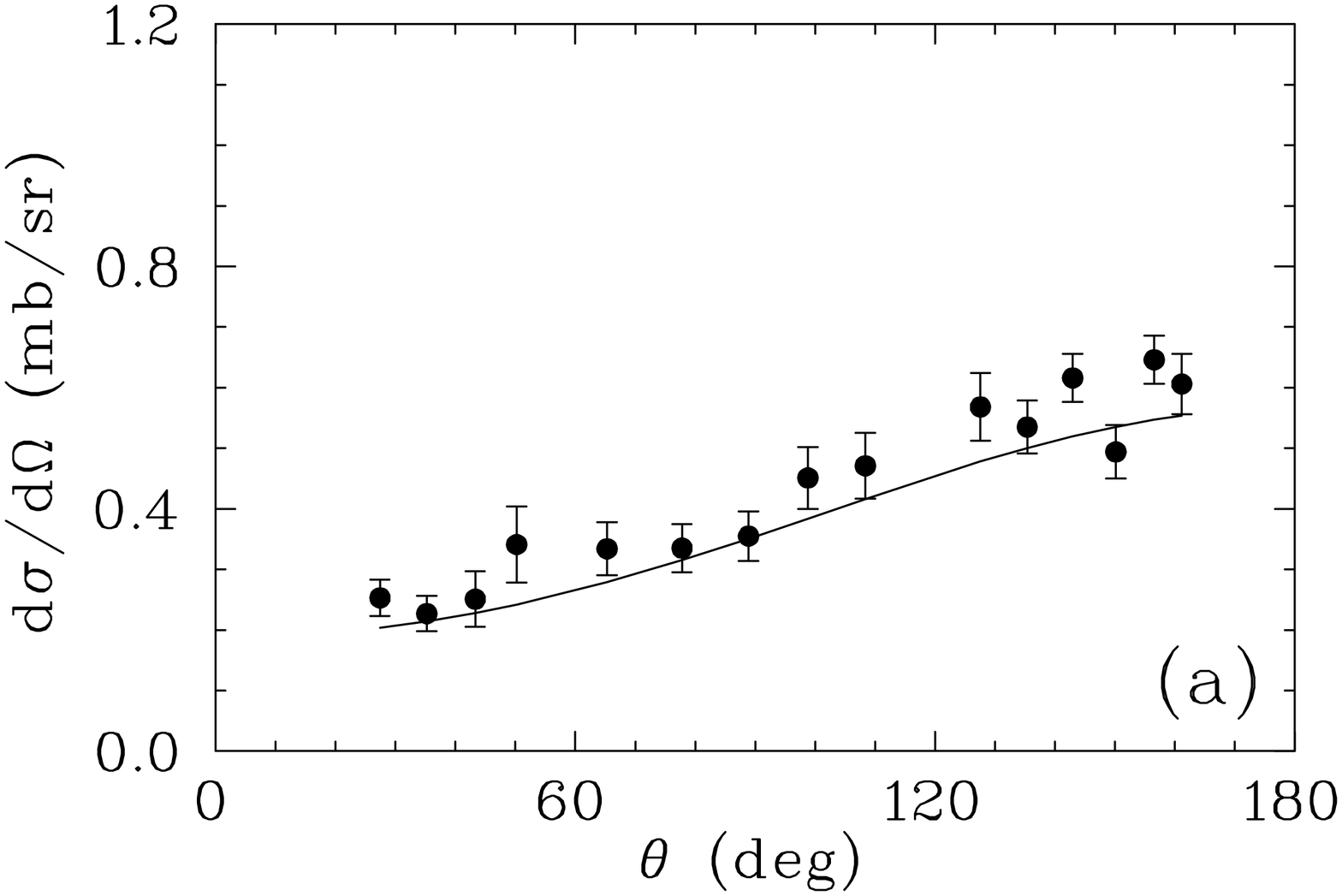}\hfill
\includegraphics[height=0.4\textwidth, angle=90]{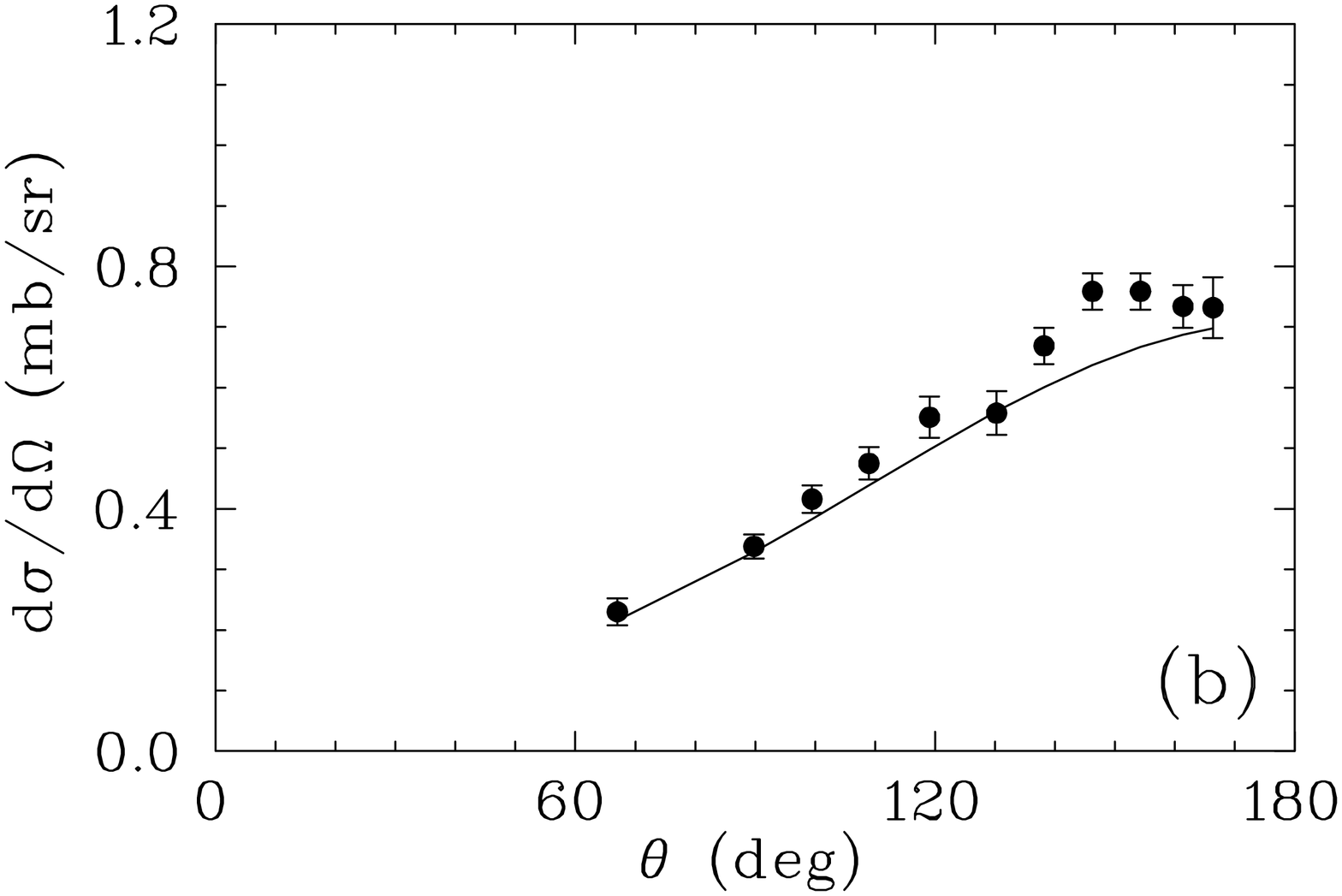}
\includegraphics[height=0.4\textwidth, angle=90]{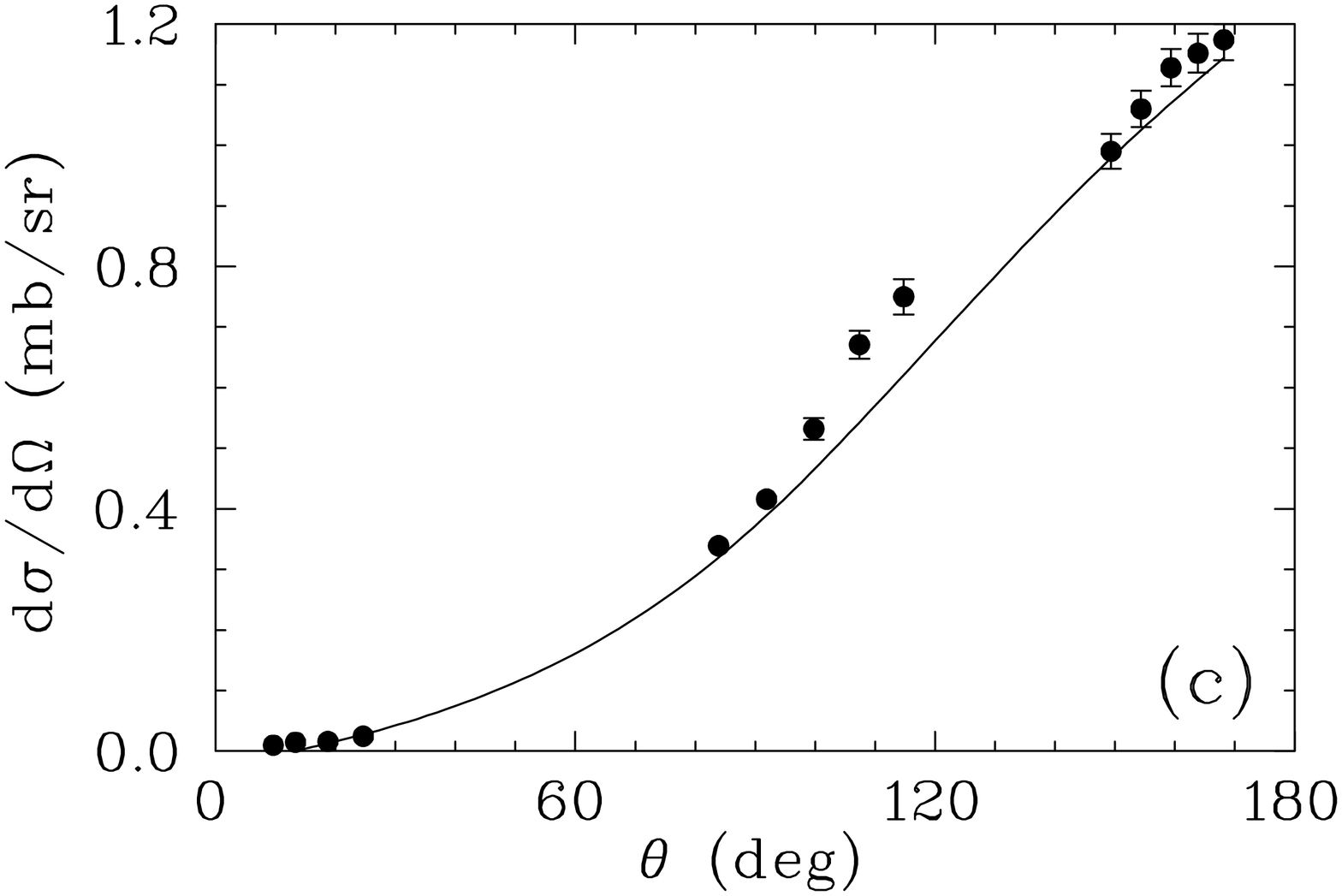}
} \caption{The differential cross sections for the
         $\pi^-p\to\pi^0n$ reaction.  (a) $T_{\pi}=
         10.6$~MeV, (b) $T_{\pi}=20.6$~MeV, and
         (c) $T_{\pi}=39.4$~MeV.  Solid line is
         the result of the present work.  Solid circles
         represent data from~\protect\cite{isenh}.
         \label{fig:g7}}
\end{figure}

\begin{figure}[ht]
\centering{
\includegraphics[height=0.4\textwidth, angle=90]{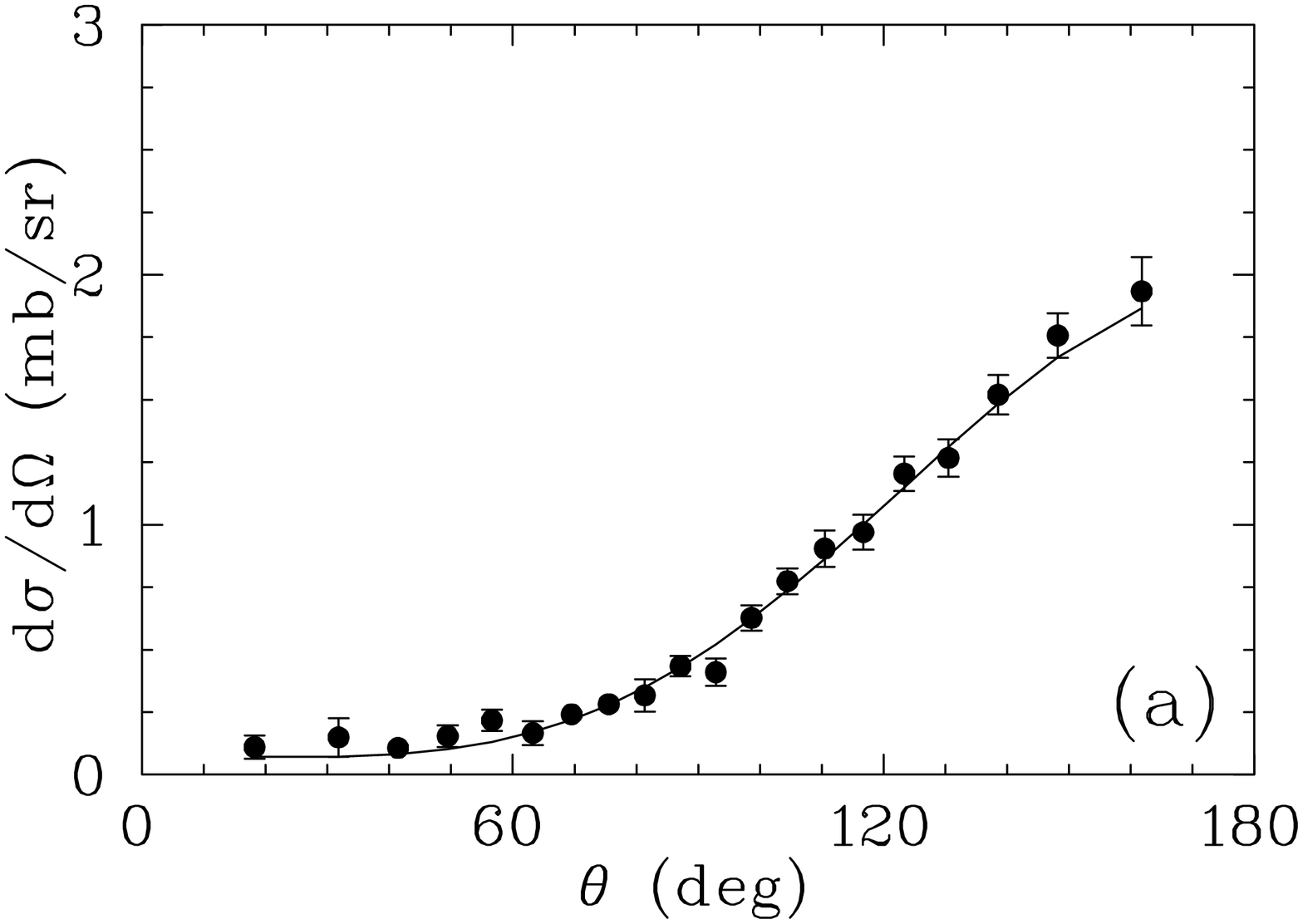}\hfill
\includegraphics[height=0.4\textwidth, angle=90]{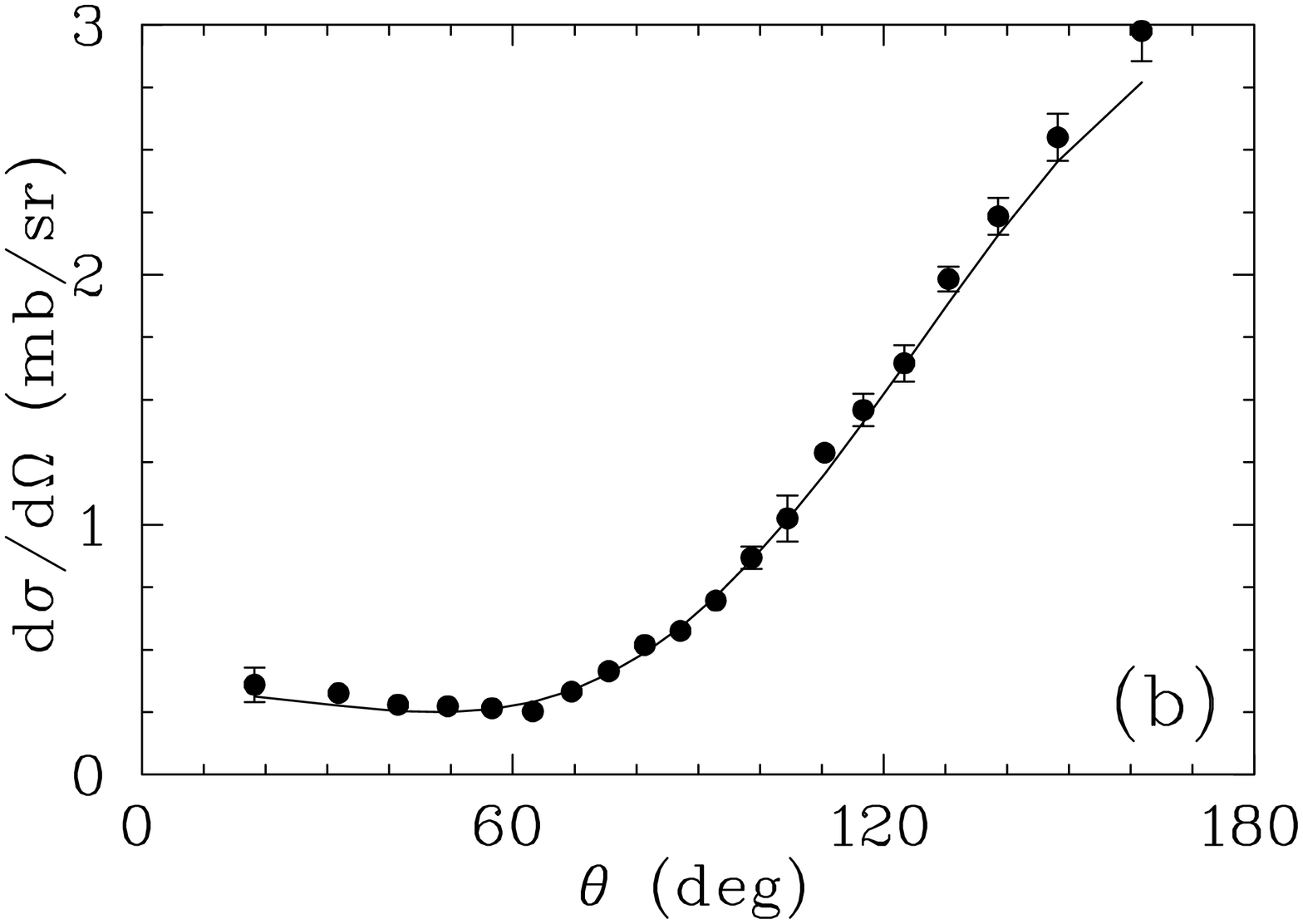}
\includegraphics[height=0.4\textwidth, angle=90]{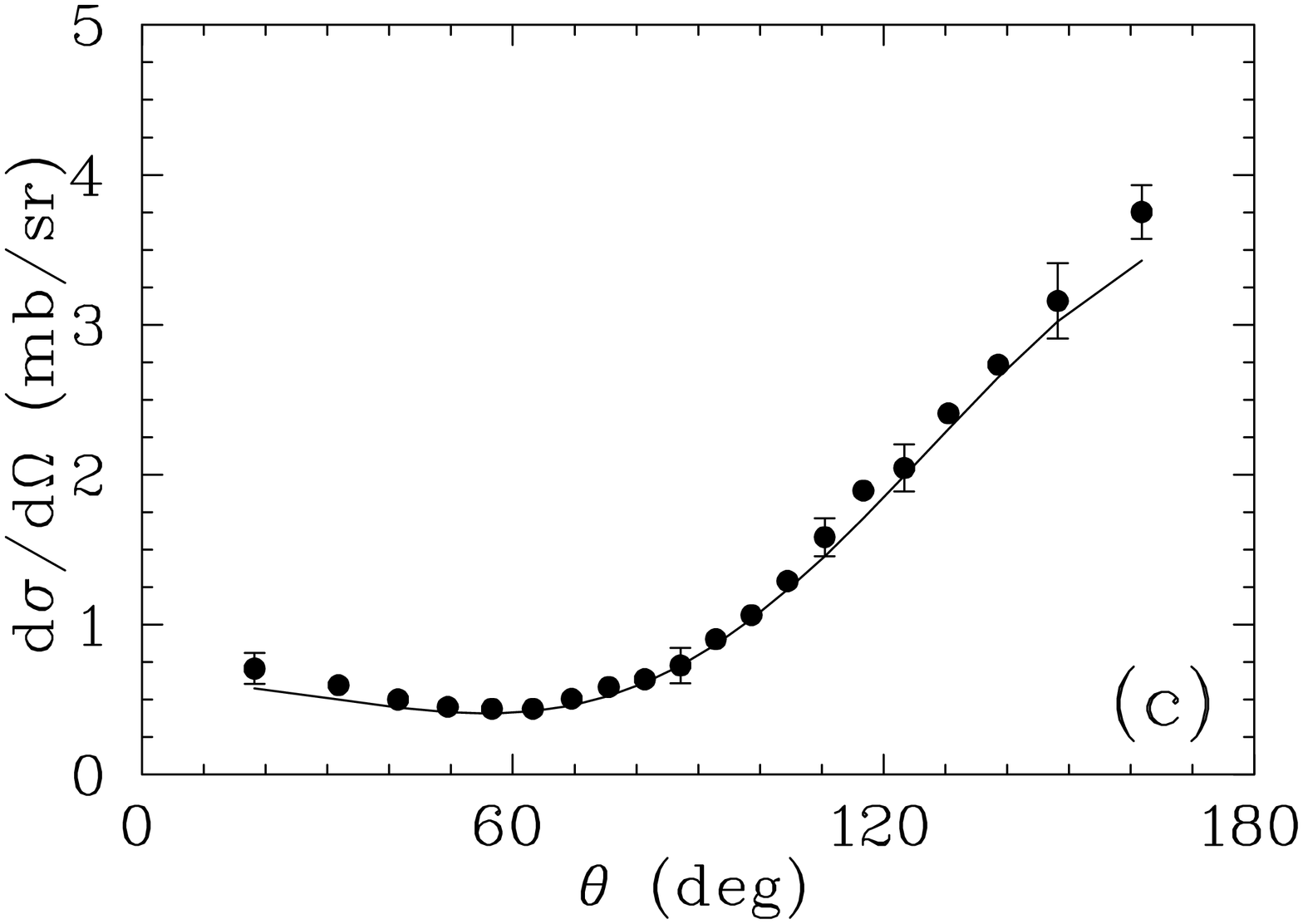}\hfill
\includegraphics[height=0.4\textwidth, angle=90]{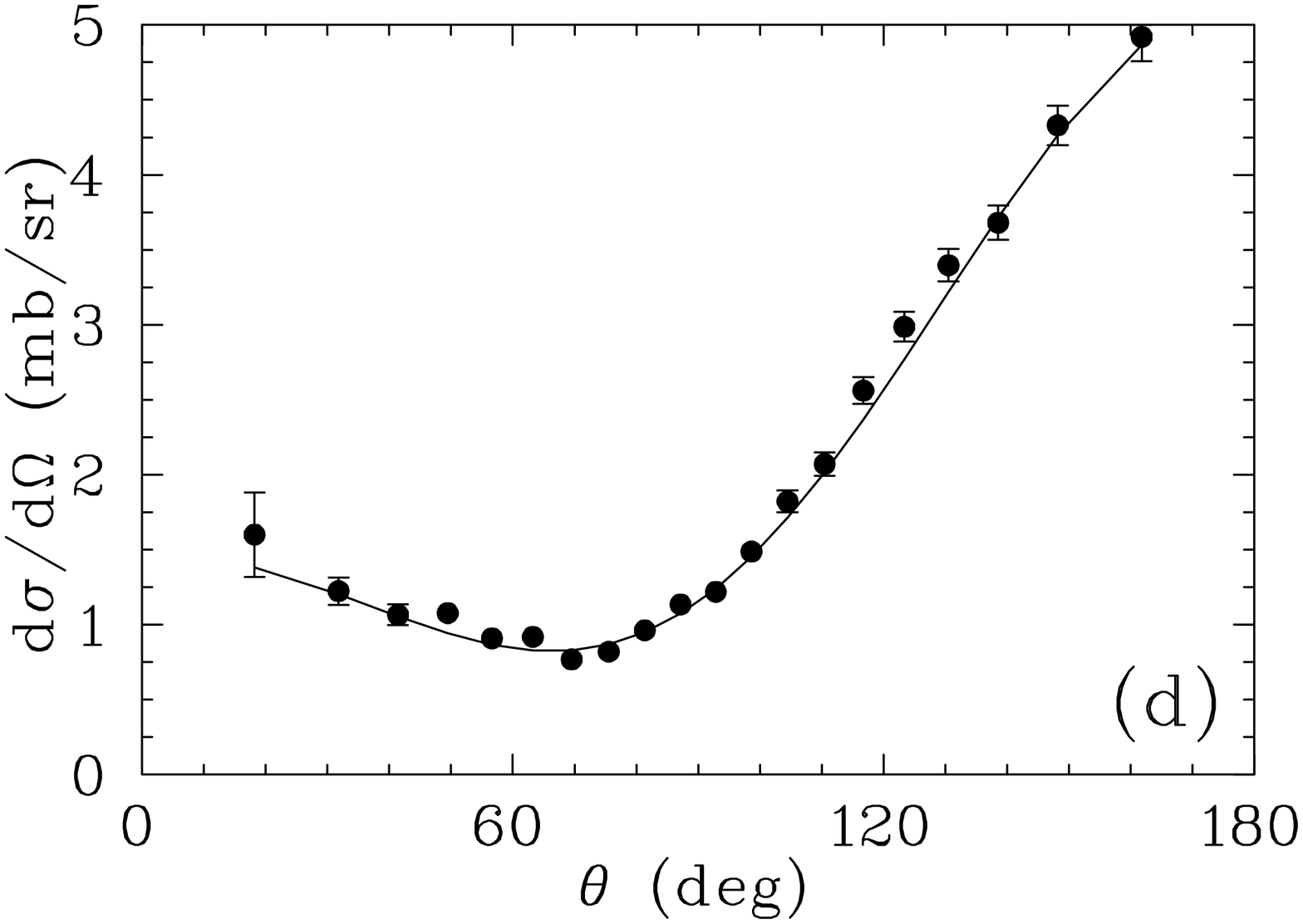}
\includegraphics[height=0.4\textwidth, angle=90]{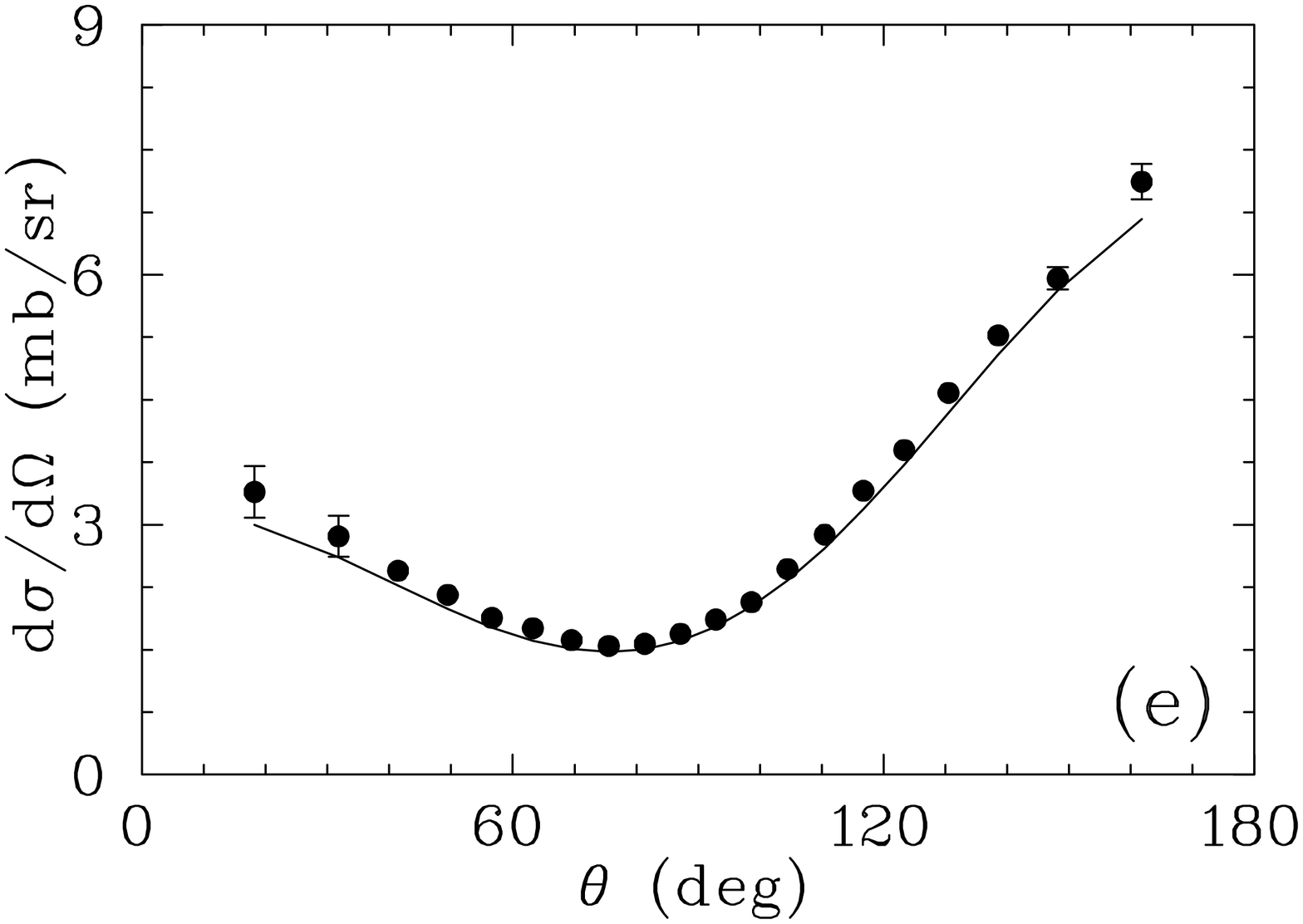}\hfill
\includegraphics[height=0.4\textwidth, angle=90]{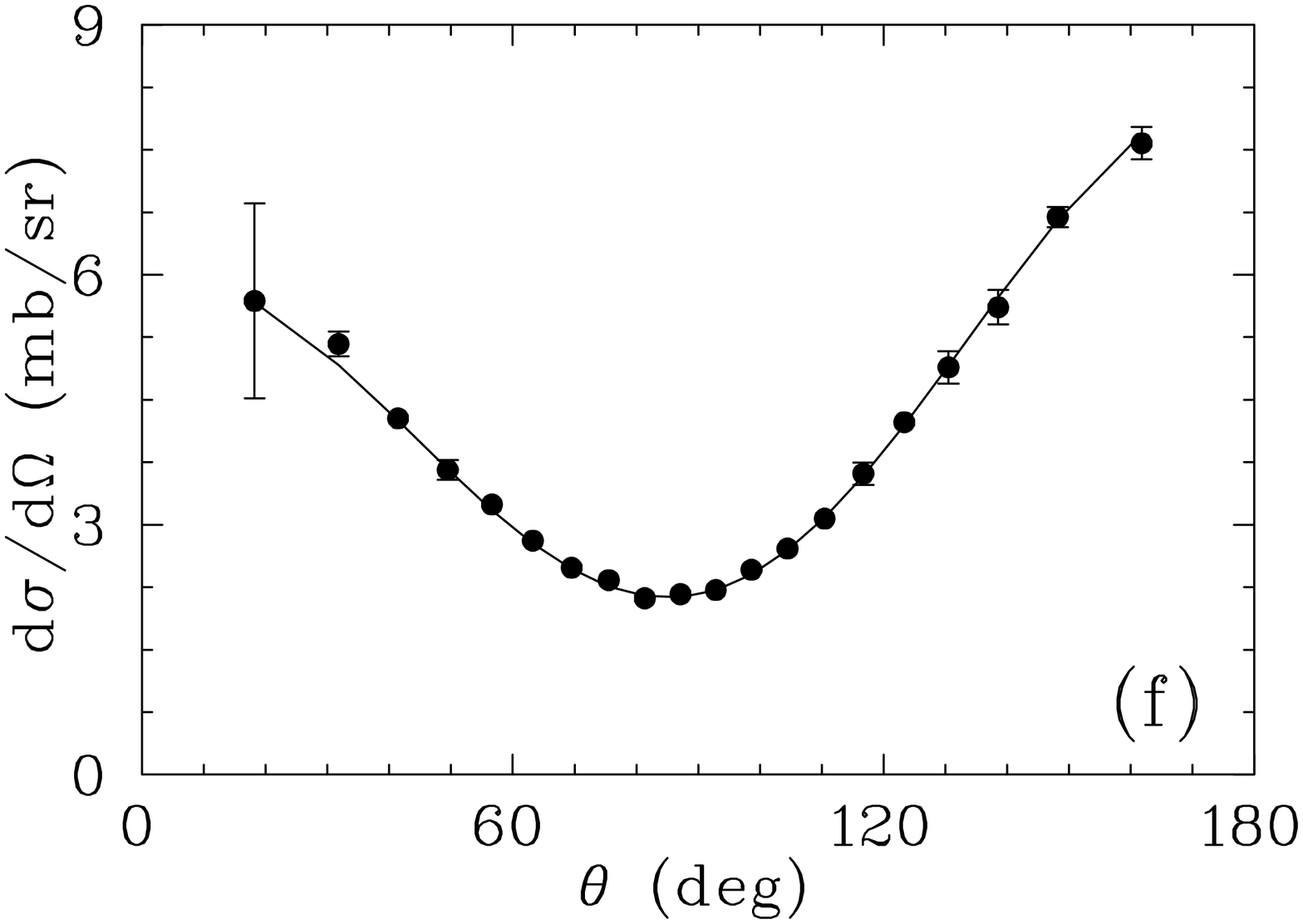}
\includegraphics[height=0.4\textwidth, angle=90]{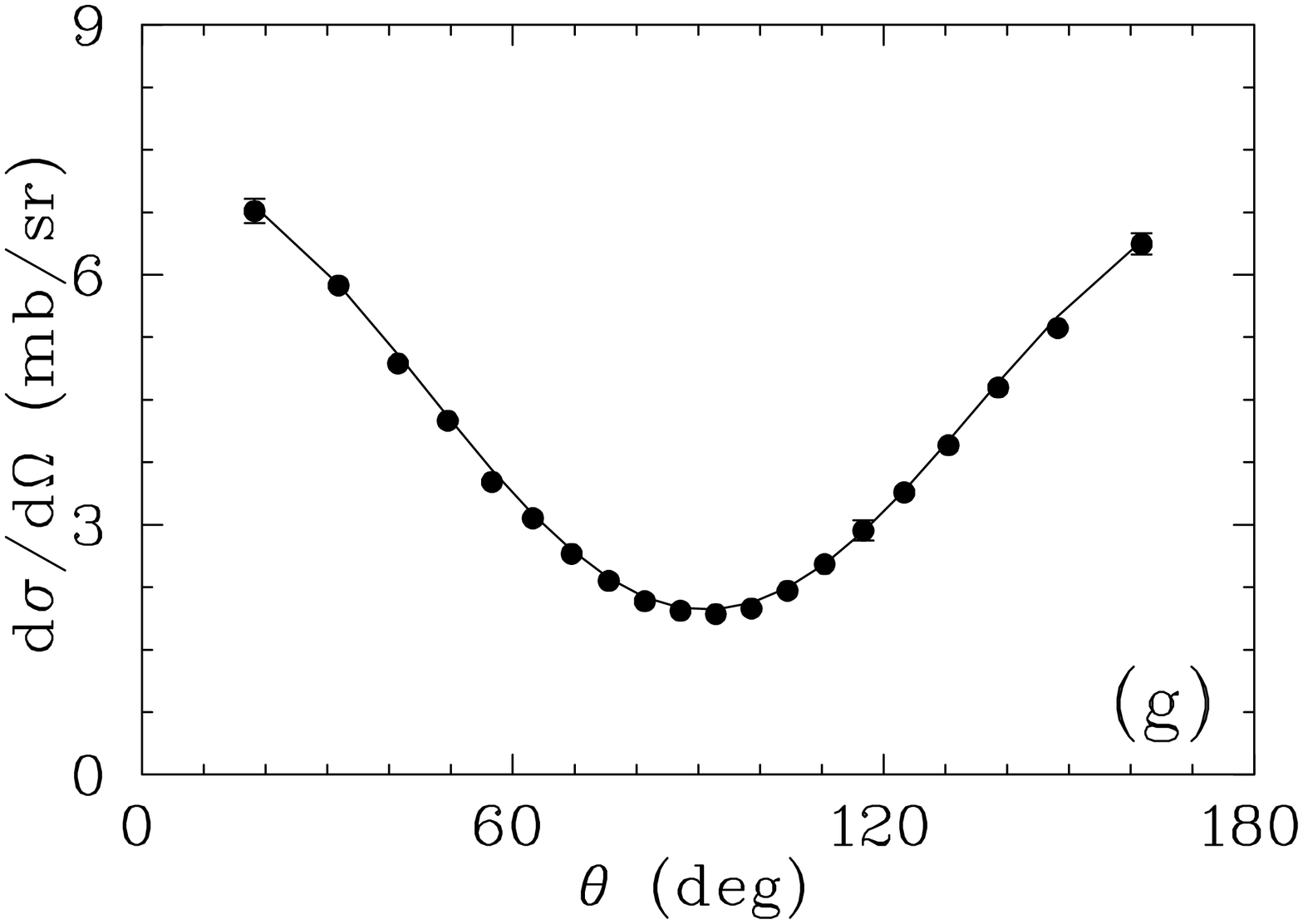}\hfill
\includegraphics[height=0.4\textwidth, angle=90]{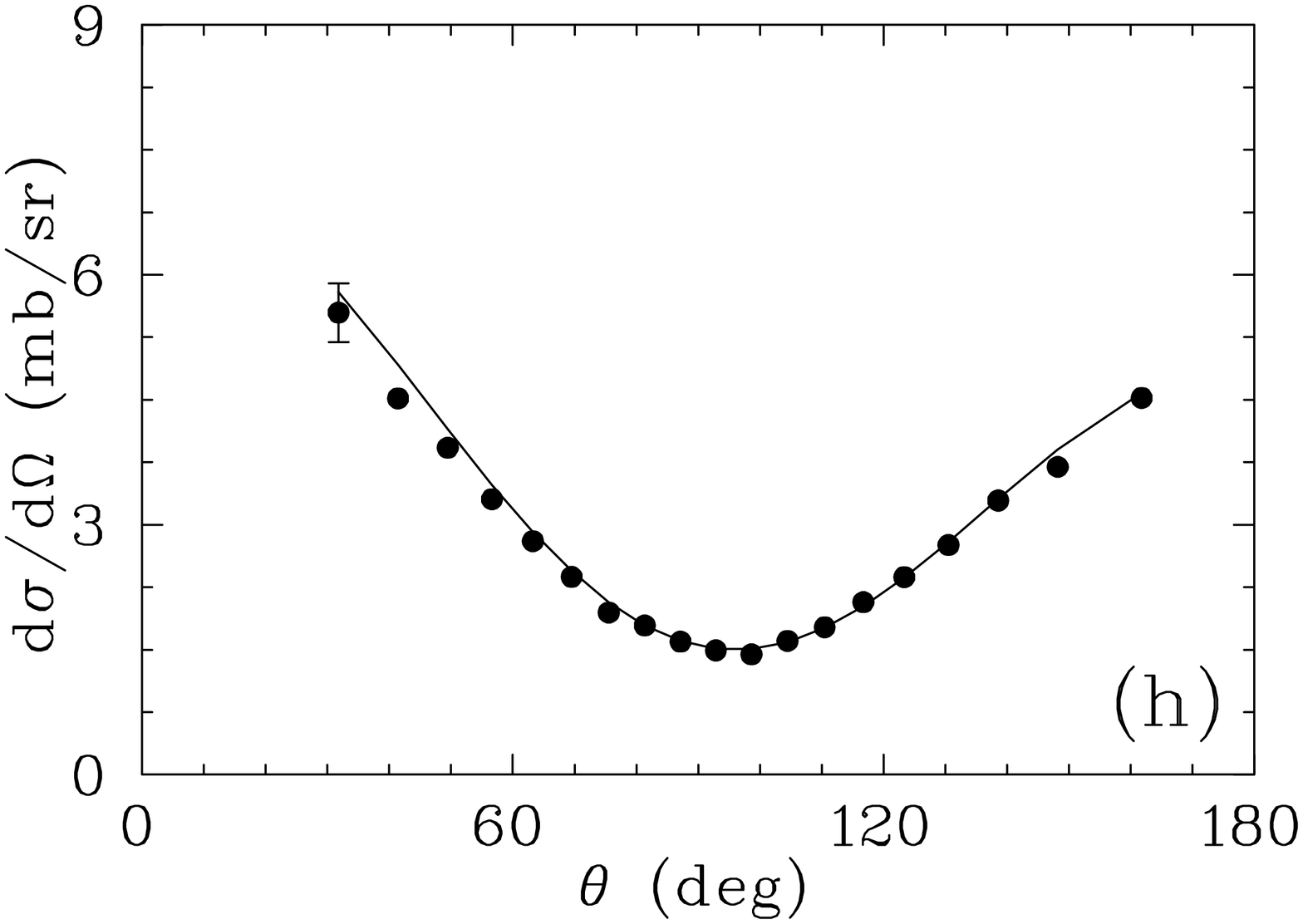}
} \caption{The differential cross sections for the
         $\pi^-p\to\pi^0n$ reaction.  (a) $T_{\pi} =
         64.1$~MeV, (b) $T_{\pi} = 83.6$~MeV, (c) $T_{\pi} =
         95.1$~MeV, (d) $T_{\pi} = 114.7$~MeV, (e) $T_{\pi} =
         136.0$~MeV, (f) $T_{\pi} = 165.6$~MeV, (g) $T_{\pi} =
         189.4$~MeV, and (h) $T_{\pi} = 212.1$~MeV.  Line
         represents predictions of the present work.
         Solid circles are recent data from
         ~\protect\cite{mike}.
         \label{fig:g8}}
\end{figure}

The CEX data play an important role in the analysis. In
Fig.~\ref{fig:g7}, we compared our results with the very low--energy
charge--exchange reaction cross section data~\cite{isenh} (these data
were included in the fit).  In Fig.~\ref{fig:g8}, the
predictions of the model are compared with the recent Crystal Ball
data taken at BNL--AGS (these data are not included in the
fit)~\cite{mike}.  The good agreement between calculations and data
is observed in both cases.

%%%%%%%%%%%%%%%%%%%%%%%%%%%%%%%%%%%%%%%%%%%%%%%%%%%%%%%%%%
\section{Conclusions}

The multi--channel tree--level $K$--matrix approach with physical
values for particle masses was developed. Isospin--conservation was
not assumed. Mass corrections to phase--shifts due to particle mass
difference were calculated and found to be in a good agreement with
NORDITA results. An isospin--symmetric version of the model leads
to reasonable agreement with the results of the latest phase--shift
analyses.  New values for $\Delta(1232)$ masses and widths were
determined directly from the experimental data. No statistically
proved sources of isospin--violation except $\Delta(1232)$ mass
difference were found. This is in a good agreement with recent
calculations based on Chiral Perturbation Theory ~\cite{meiss,fett} .
Coupling constants $g_{\pi N \Delta}$ for all charged channels were
found to be equal within 0.2\%.  A very good agreement with
low--energy CEX data~\cite{isenh} and recent Crystal Ball
collaboration~\cite{mike} data was observed.

%%%%%%%%%%%%%%%%%%%%%%%%%%%%%%%%%%%%%%%%%%%%%%%%%%%%%%%%%%%%
\section{Acknowledgements}

We thank A.~E.~Kudryavtsev for valuable discussions.
We thank The George Washington University Center for Nuclear
Studies and Research Enhancement Fund, and the U.S. Department of
Energy for their support. One of authors (A.G.) also would like
to thank H.~J.~Leisi and E.~Matsinos for many useful discussions.

%%%%%%%%%%%%%%%%%%%%%%%%%%%%%%%%%%%%%%%%%%%%%%%%%%%%%%%%%%%%

%%%%%%%%%%%%%%%%%%%%%%%%%%%%%%%%%%%%%%%%%%%%%%%%%%%%%%%%
\newpage
%%%%%%%%%%%%%%%%%%%%%%%%%%%%%%%%%%%%%%%%%%%%%%%tbl1
\begin{table}[th]
\caption{Parameters of the model (the $G_\rho^V$  and
         $G_{\sigma\pi}$ are given in GeV$^{-2}$).
         \label{tbl1}}
\begin{tabular}{|l|c|}
\colrule
$G_{\rho}^{V}$        &  44.7   $\pm$   3.0          \\
$G_{\sigma\pi}$       &  24.5   $\pm$   0.7          \\
$\kappa$              &  1.9    $\pm$   0.40         \\
$g^{2}_{\pi NN}/4\pi$ &  13.8   $\pm$   0.1          \\
$x_{\pi N}$           &  0.05   $\pm$   0.01         \\
$g_{\pi N \Delta}$    & 28.91   $\pm$   0.07         \\
$Z_{\Delta}$          & -0.332  $\pm$   0.008        \\
\colrule
\end{tabular}
\end{table}
%%%%%%%%%%%%%%%%%%%%%%%%%%%%%%%%%%%%%%%%%%%%%%%%%%%tbl2
\begin{table}[th]
\caption{Masses and widths of $\Delta(1232)$ isobar
         (all quantities are given in MeV). \label{tbl2}}
\begin{tabular}{|l|c|c|c|c|c|c|}
\colrule
& $M_{\Delta^{++}}$& $M_{\Delta^0} $& $ M_{\Delta^0}-M_{\Delta^{++}} $&
$\Gamma_{\Delta^{++}}$ &$\Gamma_{\Delta^{0}}$
&$\Gamma_{\Delta^{0}}-\Gamma_{\Delta^{++}}$        \\
\colrule
 Present work                   & 1230.55 $\pm$ 0.20 & 1233.40 $\pm$ 0.22 &
2.86 $\pm$ 0.30 &112.2 $\pm $0.7 & 116.9 $\pm $0.7 & 4.66$\pm $ 1.00 \\
 Koch \textit{et al.}~\protect\cite{koch}
                                & 1230.9  $\pm$ 0.3  & 1233.6  $\pm$ 0.5  &
2.70 $\pm$ 0.38 &111.0 $\pm $1.0 & 113.0 $\pm $1.5 & 2.0 $\pm $ 1.0 \\
 Pedroni \textit{et al.}~\protect\cite{pedroni}
                                & 1231.1  $\pm$ 0.2  & 1233.8  $\pm$ 0.2  &
2.70 $\pm$ 0.38 &111.3 $\pm $0.5 & 117.9 $\pm $0.9 & 6.6 $\pm $ 1.0 \\
 Abaev \textit{et al.}~\protect\cite{abav}
                                & 1230.5  $\pm$ 0.3  & 1233.1  $\pm$ 0.2  &
2.6  $\pm$ 0.4  &                &                 & 5.1 $\pm $ 1.0    \\
 Arndt \textit{et al.}~\protect\cite{fa02}
                                &                    &                    &
1.74 $\pm$ 0.15 &                &                 & 1.09$\pm $ 0.64    \\
 Bugg1~\protect\cite{bugg}      & 1231.45 $\pm$ 0.30 & 1233.6  $\pm$ 0.3  &
1.86 $\pm$ 0.40 &114.8 $\pm $0.9 & 116.4 $\pm $0.9 & 1.6 $\pm $ 1.3  \\
 Bugg2~\protect\cite{bugg}      & 1231.0  $\pm$ 0.3  & 1232.85 $\pm$ 0.30 &
2.16 $\pm$ 0.40 &115.0 $\pm $0.9 & 118.3 $\pm $0.9 & 3.3 $\pm $ 1.3  \\
\colrule
\end{tabular}
\end{table}

%%%%%%%%%%%%%%%%%%%%%%%%%%%%%%%%%%%%%%%%%%%%%%%%%%%%%%
\newpage
% === PSFIG 1 ======================================
% === PSFIG 2 ======================================
% === PSFIG 3 ======================================
\newpage
% === PSFIG 4 ======================================
% === PSFIG 5 ======================================
% === PSFIG 6 ======================================
% === PSFIG 7 ======================================
% === PSFIG 8 ======================================
%%%%%%%%%%%%%%%%%%%%%%%%%%%%%%%%%%%%%%%%%%%%%%%%%%%%%%%%%

\clearpage
       А. Б. Гриднев, И. Хорн, В. Д. Бриски, И. И. Страковский\\
       \\
      $ K$-матричный подход к расщеплению массы $\Delta -$ резонанса\\
       и нарушению изоспина в $\pi N$ рассеянии при малых энергиях.\\
\\
Экспериментальные данные по $\pi N$--рассеянию в упругой области
энергий $T_{\pi} \le$ 250~МэВ анализируются в рамках многоканального
$K$--матричного подхода с эффективными лагранжианами. В данном анализе
не предполагается изоспиновая инвариантность и для масс частиц
используются их наблюдаемые значения. Вычисленные поправки за счет
разности масс $\pi^+-\pi^0$ и $p-n$ хорошо согласуются с результатами
NORDITA. Получено хорошее описание всех экспериментальных наблюдаемых.
Из анализа  данных определены новые значения для масс и ширин $\Delta^0$--
and $\Delta^{++}$--резонансов.  Фазовые сдвиги $\pi N$--рассеяния в
изоспиново--симметричном случае близки к новому решению FA02 фазового
анализа, полученному в Университете Джорджа Вашингтона, и основанному
на самых современных экспериментальных данных. Наш анализ приводит
к значительно меньшему ($\le $1\%) нарушения изоспина в интервале
энергий $T_{\pi}\sim 30-70$~МэВ, чем 7\% , полученное в работах
Gibbs  \textit{и др.} и Matsinos, и согласуются с результатами
вычислений, основанных на киральной теории возмущений.
\end{document}